\documentclass[journal]{IEEEtran}
\usepackage{cite}

 \usepackage[pdftex]{graphicx}
\usepackage{caption}
\usepackage{subcaption}
\ifCLASSINFOpdf
\else
\fi
\usepackage{amsmath,amsfonts,bm}

\def\eqref#1{equation~\ref{#1}}

\def\1{\bm{1}}

\def\vtheta{{\bm{\theta}}}

\def\vb{{\bm{b}}}

\def\vl{{\bm{l}}}

\def\vt{{\bm{t}}}
\def\vu{{\bm{u}}}

\def\vx{{\bm{x}}}
\def\vy{{\bm{y}}}
\def\vz{{\bm{z}}}

\DeclareMathAlphabet{\mathsfit}{\encodingdefault}{\sfdefault}{m}{sl}
\SetMathAlphabet{\mathsfit}{bold}{\encodingdefault}{\sfdefault}{bx}{n}

\usepackage{url}

\usepackage{color}
\usepackage{hyperref}

\begin{document}
\bstctlcite{IEEEexample:BSTcontrol}
%
\title{A Deep Learning Approach for Digital Color Reconstruction of Lenticular Films}
%
%
%

\author{Stefano~D'Aronco, 
        Giorgio~Trumpy, 
        David~Pfluger, 
        and~Jan~Dirk~Wegner
        }
\maketitle

\begin{abstract}
We propose the first accurate digitization and color reconstruction process for historical lenticular film that is robust to artifacts. Lenticular films emerged in the 1920s and were one of the first technologies that permitted to capture full color information in motion. The technology leverages an RGB filter and cylindrical lenticules embossed on the film surface to encode the color in the horizontal spatial dimension of the image. To project the pictures the encoding process was reversed using an appropriate analog device. In this work, we introduce an automated, fully digital pipeline to process the scan of lenticular films and colorize the image.
Our method merges deep learning with a model-based approach in order to maximize the performance while making sure that the reconstructed colored images truthfully match the encoded color information. Our model employs different strategies to achieve an effective color reconstruction, in particular (\emph{i}) we use data augmentation to create a robust lenticule segmentation network, (\emph{ii}) we fit the lenticules raster prediction to obtain a precise vectorial lenticule localization, and (\emph{iii}) we train a colorization network that predicts interpolation coefficients in order to obtain a truthful colorization.
We validate the proposed method on a lenticular film dataset  and compare it to other approaches.
Since no colored groundtruth is available as reference, we conduct a user study to validate our method in a subjective manner. The results of the study show that the proposed method is largely preferred with respect to other existing and baseline methods.
\end{abstract}


%
\IEEEpeerreviewmaketitle

%
%
%
%

\section{Introduction}

Historical films represent a precious source of information for how the world and life looked like in the past. Film digitization is essential to protect these valuable records from deterioration and also to facilitate their visualization. 
Digitization and visualization of historical films is not always a trivial process, as, in some cases, these media might rely on a specific analog device for their visualization. As a result, a software adaptation of the original analog process is needed to display them today.

\begin{figure}
     \centering
     \begin{subfigure}[c]{\columnwidth}
         \centering
         \includegraphics[height=3cm]{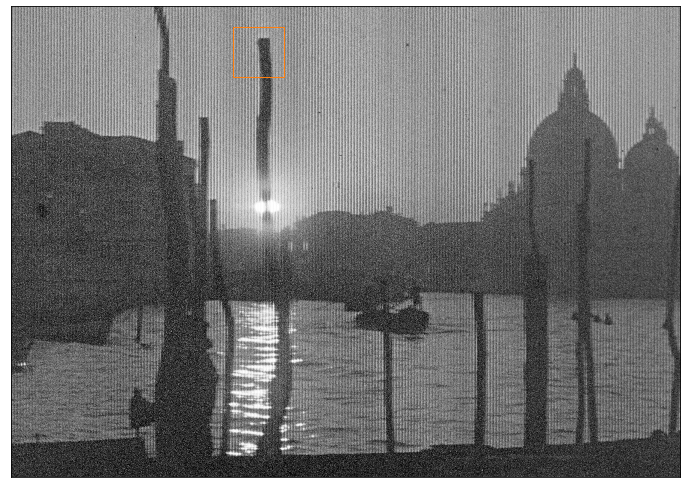}
         \ \ \
         \includegraphics[height=3cm]{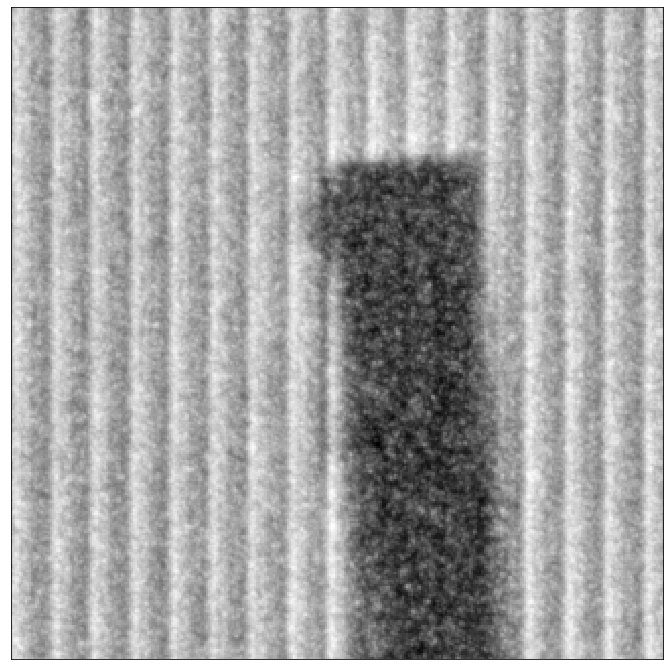}
    \end{subfigure}\\
    \vspace{0.2mm}
    \begin{subfigure}[c]{\columnwidth}
         \centering
         \includegraphics[height=3cm]{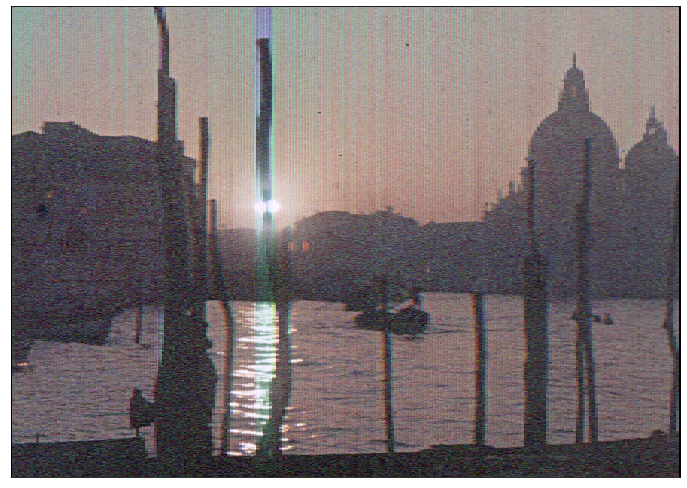}
         \ \ \ 
         \includegraphics[height=3cm]{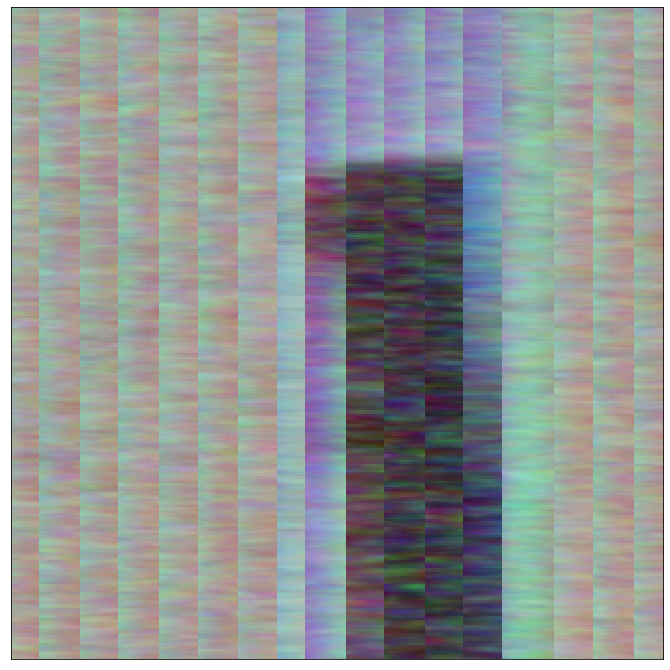}
     \end{subfigure}\\
    \vspace{0.2mm}
     \begin{subfigure}[c]{\columnwidth}
         \centering
         \includegraphics[height=3cm]{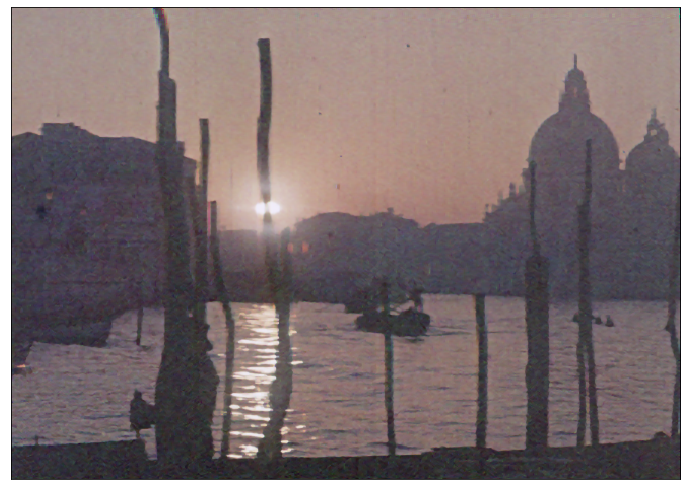}
         \ \ \ 
         \includegraphics[height=3cm]{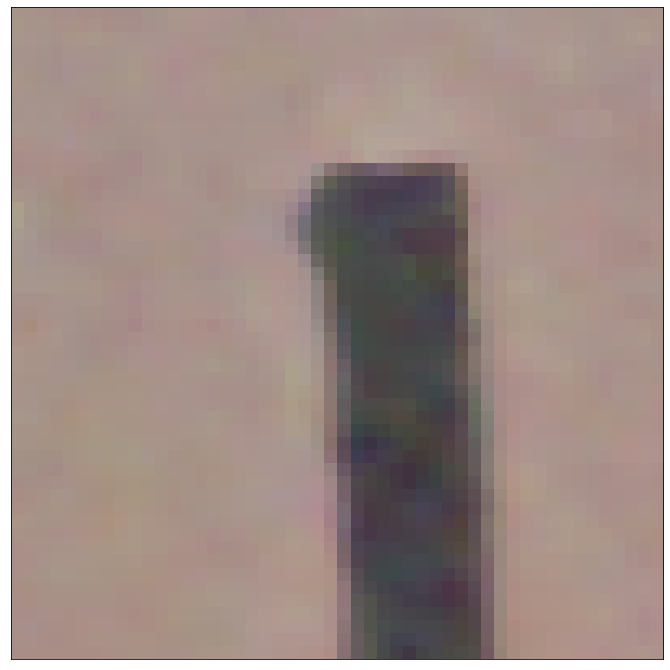}
     \end{subfigure}
        \caption{Example of a scanned lenticular film (top), and doLCE color reconstruction (center), proposed method (bottom)}
        \label{fig:example_colorized}
\end{figure}

Lenticular film, made popular by Eastman Kodak  under the  name \emph{Kodacolor} in the late 20s, represents one of the earliest technologies that allowed amateur filmmakers to record color movies~\cite{gordon2013}. The technology consisted of a black-and-white film embossed with vertical cylinders, called lenticules, and an RGB filter placed in front of the camera lens. This combination allowed to encode the color information as silver densities in the horizontal spatial dimension within each lenticule. %
After being exposed in the camera, the film underwent reversal processing.
Its visualization was done with a dedicated projection device that reversed the recording process, using an RGB filter in front of the lens to create color images from the black-and-white film.
Different approaches can be adopted to recreate the color images of lenticular film with digital tools. The digitization of the projected, colored film is time consuming and requires a special, obsolete projection device. Scanning the black-and-white lenticular film, on the other hand, is definitely more practical, but the additive colors created optically during analog projection have to be created numerically via software. 
Two main steps are required to colorize a scanned lenticular film: \emph{i}) identify the lenticules on the film, and \emph{ii}) extract the grayscale values inside each lenticule to finally reconstruct the RGB image. 
The lenticules on the scanned film can be detected by looking at their boundaries which appear like dark vertical straight lines. After the boundaries of a lenticule are identified, the RGB information, which is stored in the horizontal spatial dimension between the boundaries, can be easily extracted. An example of a scanned input film, grayscale, with the color reconstruction is shown in Fig.~\ref{fig:example_colorized}.

The appropriate color reconstruction strategy for a scanned, lenticular film depends on the final, intended application. For instance, the principle dear to the film preservation community is to obtain the closest possible reproduction to the original, analog projection process, accurately simulating how the images would look like when visualized with the historical, analog projector. Another less orthodox aim would be creating images that simulate how the scenes represented in the lenticular film would look if they were shot with modern cameras by hallucinating some details and altering the colors.
We aim to obtain a digital version of the lenticular film that somewhat improves the visualization w.r.t. the analog one, while maintaining the truthfulness dictated by the resolution and color information encoded in the lenticular film. In particular, we do not aim to suppress specific artifacts like scratches or other large defects potentially present on the lenticular film in order to fully preserve the historical value of the pictures. At the same time we aim to build a network that is more robust w.r.t. existing digital lenticular color reconstruction algorithms, and avoid severe artifacts that arise from an erroneous lenticule localization.

We design a multistage pipeline that combines deep learning and model-based stages. The first stage is a segmentation network that performs a binary classification on the input image, discriminating between pixels located on lenticules' boundaries and pixels within the lenticules. We then leverage this prediction in order to extract the lenticules' boundaries in a vector format, rather than a raster prediction. For each detected lenticule, the RGB values encoded are extracted creating an image where for each pixel column only one of the RGB channels is known. A second CNN, trained on a separate dataset, is then used in order to interpolate between the missing color values to colorize the image in what resembles a \emph{demosaicing} step. 
To guarantee a more truthful color reconstruction the network is constrained to predict the output color by interpolating between nearby pixels. 
Note that in this case we do not have access to a dataset made of pairs of scanned lenticular films and colorized images as ground truth, as a result we need to use a secondary dataset to train the colorization network.
To the best of our knowledge the proposed method is the first learning based color reconstruction method for lenticular films. 
We validate the proposed pipeline on a dataset of scanned lenticular films and show that our method effectively colorizes the images while reducing some artifacts which appear with basic digital colorization techniques.
We further conduct a user study to have a subjective evaluation of different color reconstruction techniques and show that the proposed method is largely preferred over other approaches.


\section{Related Work}\label{sec:related}

Much literature exists on historical photo restoration~\cite{survey_restoration}. Some early works mainly focused on identifying scratches on the photos using some hand crafted features~\cite{identify_scratches,identify_scratches2}. Removing scratches that appear on photos, especially when caused by \emph{mechanical} damage, is rather complicated as those defects are usually large. As a result, an inpainting method is needed to fill the voids in the picture after removing the damaged area, a field of research where deep learning has proven to be very effective~\cite{deep_inpaint1,deep_inpaint2}.
The authors in~\cite{back_to_life} propose a deep learning method for the restoration of historical photos. Their approach is composed of three variational autoencoders and a translational network. The main problem when training a network for historical photo restoration is that no real pairs of degraded and restored images exist. In~\cite{back_to_life} degraded images are generated artificially so that it is possible to learn a mapping to reconstruct the restored images. To make sure that the reconstruction pipeline works on old images, the latent space of the artificially degraded images and the one of the old damaged photos are forced to be aligned. The reconstruction loss is composed of a mix of $l_1$ loss, adversarial loss, and perceptual loss. The same authors recently proposed a similar but more advanced network with a face enhancement module~\cite{back_to_life_jrnl}.

Colorization is a more general problem that is not exclusively tailored to old photos restoration. A survey on different colorization approaches and applications can be found in~\cite{survey_colorization}. In the simplest colorization settings, the aim is to reconstruct the RGB triplets starting from the single luminance component. Some details need to be hallucinated because many real-life objects exist in a wide range of colors. An effective pipeline for colorization of historical images is the \emph{DeOldify} project~\cite{deoldify}. DeOldify is able to colorize single images or videos, starting from a grayscale image, and uses a Generative Adversarial Network (GAN) in order to obtain realistic-looking, colorized images.

Most of the works about colorization, however, are not suited to deal with lenticular films because they are not designed to leverage the color information encoded in the spatial domain. If we assume that the lenticules are detected and the corresponding RGB values are extracted, the color reconstruction process of lenticular films shares a lot of similarities with demosaicing~\cite{old_demosaic,mosaic_survey}. The problem of demosaicing consists in reconstructing a full color image starting from an incomplete image, where only a single value among red, green, and blue is known for each pixel. This problem usually originates from the usage of an array color filters in front of an image sensor that creates a mosaic structure among the sensed color channels. In the case of lenticular film, the structure is composed of vertical stripes rather then a mosaic. 
Basic algorithms for demosaicing use rather simple interpolation techniques~\cite{mosaic_survey}. \cite{deep_demosaic_1} is one of the first works that tackled the demosaicing problem using a deep learning method. Authors in~\cite{deep_demosaic_2} show that demosaicing can be seen as a particular case of super-resolution and they build a deep learning pipeline that tackles denoising, demosaicing and superresolution all at the same time with and end-to-end trainable architecture. In~\cite{deep_demosaic_attention} authors propose  an attention-based network for image restoration with demosaicing being one of the target applications. 
Previous works demonstrate the ability of deep learning to deal with problems such as demosaicing and denoising. However, none of the architectures is designed for reconstructing the color of lenticular films. 

Besides a few unpublished solutions developed by service providers, the only documented method for lenticular film is the doLCE algorithm~\cite{Reuteler2014}. The doLCE method, similarly to ours, is composed of two stages: detect the lenticules, and colorize the image. The method, although effective on many images, suffers from two main limitations: \emph{i}) the lenticule detection only works with, almost, perfectly vertical lenticules, \emph{ii}) the colorization process assigns a single RGB triplet for the entire width of the lenticule thus drastically reducing the resolution in the horizontal direction. 
In this paper we aim at addressing those limitations to design a pipeline that can deal with tilted lenticules while at the same time increasing the detail level of the output image. We build upon our preliminary work~\cite{deepdolce_2021} where we tested the feasibility of using a convolutional neural network to detect the lenticule boundaries. In our new work we include a refinement module for the lenticules extraction as well as a colorization network which preserves a higher level of detail from the grayscale input image.

\section{Background}\label{sec:background}

\begin{figure}[t]
    \centering
    \includegraphics[width=\columnwidth]{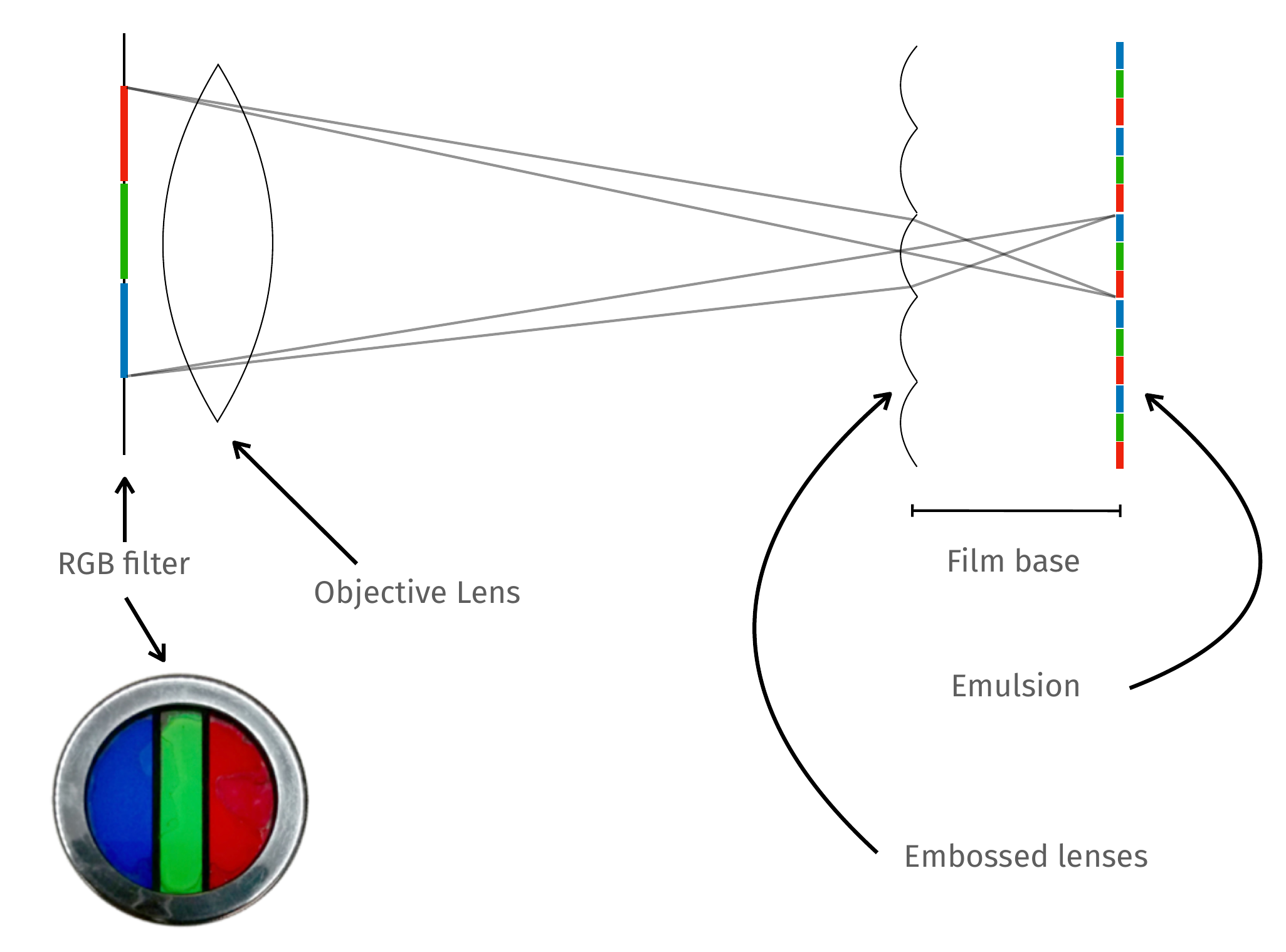}
    \caption{Side view section of the lenticular film camera. Tiny cylinders are embossed focusing the light colored on different parts of the emulsion.}
    \label{fig:lenticular_camera}
\end{figure}

\begin{figure}
     \centering
     \begin{subfigure}[c]{0.48\columnwidth}
         \centering
    \includegraphics[height=3.5cm]{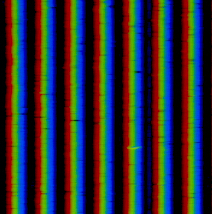}
         \caption{}
     \end{subfigure}
     \begin{subfigure}[c]{0.48\columnwidth}
         \centering
      \includegraphics[height=3.5cm]{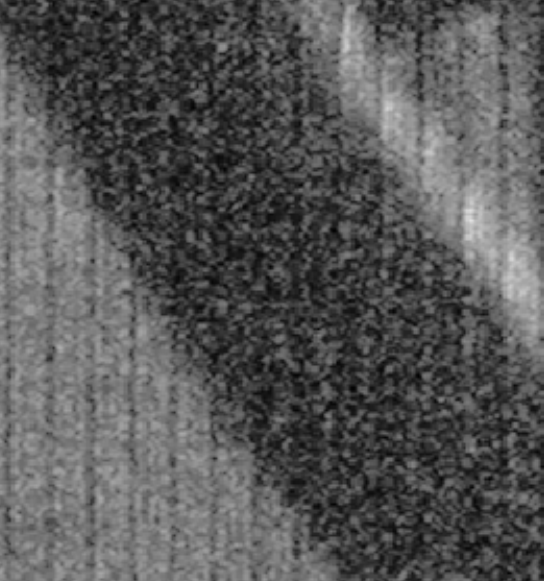}
         \caption{}
     \end{subfigure}
        \caption{Left: Position of the red, green and blue components within the lenticules. Right: detail of a diagonal edge on a scanned lenticular film. As it can be seen the diagonal edge is still perceivable within the lenticules meaning that }
        \label{fig:ex_input}
\end{figure}

The schematic diagram of a lenticular film camera is shown in Fig.~\ref{fig:lenticular_camera}. First, an RGB filter is placed in front of the camera lens. The lenticular film is characterized by a vertical array of hundreds of cylindrical lenses. The lenticules’ focal length is three times the curvature radius and is equal to the film thickness. Focal length and aperture of the camera were designed so that the photographic emulsion is exposed to the different color components~\cite{Capstaff}. 

Lenticular film was sold in 16 mm format, where the images occupies about 10 mm of film width. Considering that the width of each lenticule is about 43 $\mu$m, approximately 230 lenticules fit within each image. In order to extract the spatially-encoded color information, the film has to be scanned with the emulsion facing the image capture system. After scanning the original, historical film, the lenticules' boundaries appear as dark straight vertical lines in the digital image. 
Estimating their width, in number of pixels, is straight-forward because we know the scanning resolution. However, the lenticule width does vary slightly within each scanned film, this is possibly caused by: optical aberrations of the digital capture,  limited manufacturing quality of the film itself, time degradation, or mechanical damage. From our dataset we empirically observed that the lenticules' width inside a single frame varies by about 5\% of the average width. 
Moreover, the film is often slightly misaligned with the scanner, which leads to the lenticules not being perfectly vertical. In this regard, we empirically observed a maximum rotation of about $1^\circ$.

In Fig.~\ref{fig:ex_input}a, we show where the colors are encoded within the lenticules. This image is obtained by shining white light through the original lens of a lenticular camera and capturing with a digital camera the light that was supposed to reach the film emulsion.
In order to reconstruct the colors of the scanned grayscale image, we need to detect the lenticules, extract the color and fill each lenticule for its entire width with the RGB values. 
One question that rises when looking at the grayscale scanned film is how the color information is exactly encoded. Do the red-green-blue values inside each row of a lenticule refer to the same point in the captured scene? Or do they refer to different points in the scene where for each point we have access to only one of the RGB value?
One way to answer this question is by carefully looking how diagonal edges look like in the grayscale images.
Fig.~\ref{fig:ex_input}b shows quite sharp diagonal edges between achromatic regions. Note that in white/gray/black regions the red-green-blue values in each lenticule should roughly always amount to the same value. 
As it can be seen, inside a single lenticule the diagonal edge is still visible, this means that valuable spatial horizontal information is contained within each lenticule. In other terms, different pixels in the horizontal direction within a lenticule map to different points in captured scene. 

Finally, in order to perform an accurate color reconstruction, we need to know how to map the color values of the lenticular film in a standard RGB space. This mapping process can be derived from the transmission spectra of the RGB filter. We refer the reader to the Appendix~\ref{app:color_est} for further information about this process.

\section{Proposed Method}\label{sec:method}

We denote the input image with $\vx$, which is a grayscale image of size $H\times W$ pixels.
The proposed pipeline is split in two stages: the first stage is responsible for extracting the lenticules boundaries, whereas the second one is responsible for reconstructing the RGB triplet inside each lenticule.

\subsection{Lenticule Boundaries Detection}

The lenticule boundaries appear like straight vertical valleys (dark ridges) in the input grayscale film scan. What makes boundary detection hard is that such valleys are only visible in parts of the image that are sufficiently bright, see for instance Fig.~\ref{fig:len_pred}b.
Moreover, lenticules are in reality often not exactly vertical but slightly misaligned because of the scanning process. 
Furthermore, the distance between lenticule boundaries, though being rather regular, it is not exactly constant across the image, preventing us from fitting a regularly spaced grid. These artifacts are major reasons for failure of doLCE~\cite{Reuteler2014}, a method built on well-established image processing tools for certain easy cases: clearly visible, almost perfectly vertical lenticule boundaries for at least some part of the image along the vertical direction.
In order to generate a large dataset with clean lenticule labels for training our convolutional neural network (CNN), validation, and testing, we use successful lenticule detection cases obtained with doLCE.  
Since one major goal of our approach is greater robustness against artifacts (where doLCE fails), we artificially create many hard cases via data augmentation.
In general, any modern network architecture for image segmentation could be used for lenticule boundary detection. Training is done by minimizing the cross-entropy loss between the ground truth and the network prediction:
\begin{equation}
     \underset{\vtheta}{\text{minimize}}\ \ \ \mathbb{E}_{\vx,\vz_\text{gt}} [\text{CE}(\vz (\vx),\vz_\text{gt})],
\end{equation}
where CE represents the pixelwise cross entropy loss, $\vz(\vx)$ is the network output ($z\in[0,1]$) and  $\vz_\text{gt}$ is the ground truth segmentation map.
As already mentioned successful doLCE reconstructions occur exclusively when the lenticule boundaries are perfectly vertical, creating a bias in the training dataset as the neural network is never exposed to tilted lenticules. To correct for this bias, we perform data augmentation by randomly rotating (maximum rotation is set to $\pm 1^\circ$) the input-output patches so that the network learns how to deal with this situation. 
We train the network with patches of $256 \times 256$ pixels taken from input images $\vx$, whereas the groundtruth is obtained from lenticule boundary predictions using the doLCE algorithm.

\begin{figure*}[t]
    \centering
    \includegraphics[width=\textwidth]{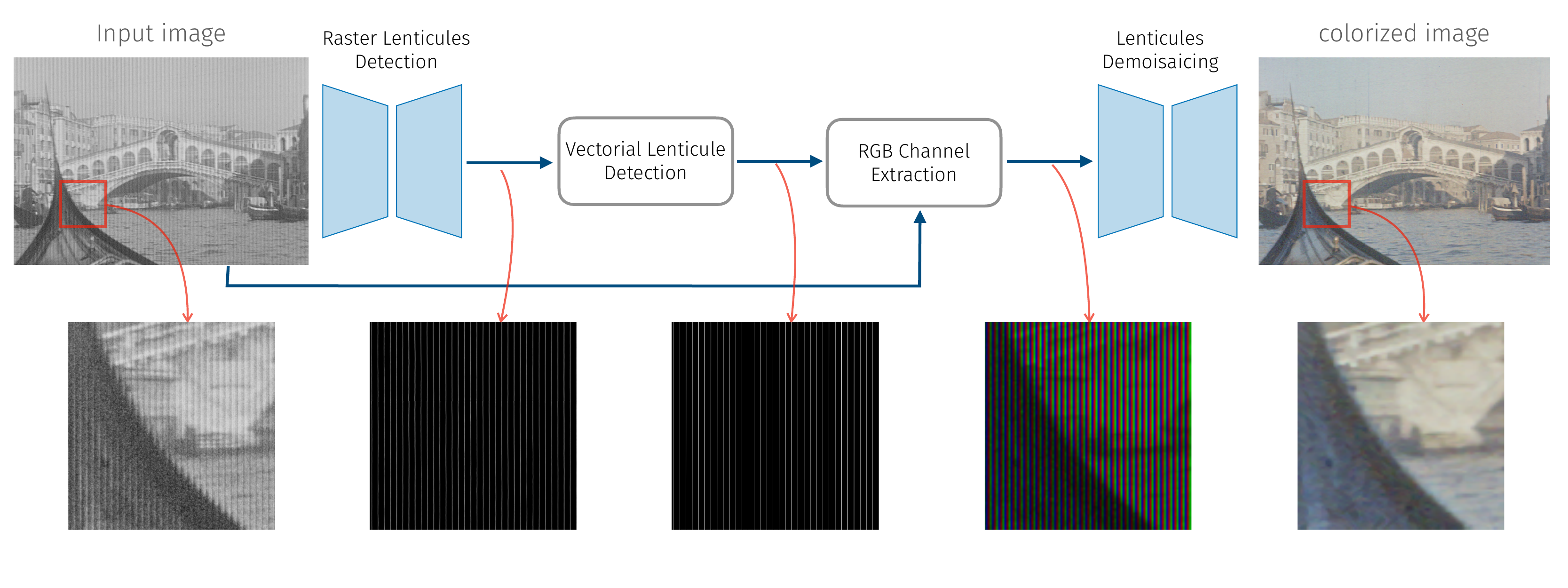}
    \caption{Overall pipeline of the proposed method. First the image is passed through a CNN for a raster lenticule boundaries detection. The prediction goes through a vectorial extraction and refinement, at this point the RGB values for each lenticule are extracted. Finally a \emph{Demosaicing} network takes care of filling the voids in the RGB channels and obtain a fully colorized image.}
    \label{fig:pipeline}
\end{figure*}

\subsection{Lenticules Refinement and Color Extraction}

The output of the neural network is a raster grayscale image that highlights the lenticule boundaries. In order to proceed with the color reconstruction, we need to extract the lenticule locations in a vector format. Thus we first define a vectorized representation of the lenticule boundaries as two sequences $\vt=(t_1,t_2,...,t_M)$ and $\vb=(b_1,b_2,...,b_M)$, where $t_n$ and $b_n$ correspond to the location (measured in pixels) of the $n$-th lenticule at the top and at the bottom of the image $\vx$. The total number of lenticule boundaries $M$ in a single image can easily be estimated from $\vz$ by simply measuring the average width of the lenticules. 
Intuitively, we need to look for the set of straight lines defined by $\vt$ and $\vb$ that best overlap with the raster lenticule boundaries predicted by the network, thus minimizing the following metric:
\begin{equation}
     \underset{\vu,\vl}{\text{minimize}}\ \ \ -\sum_{m=0}^M \sum_{h=0}^H z(h,b_m + \frac{h t_m}{H}).
     \label{eq:lent_ext1}
\end{equation}
The objective of the minimization problem corresponds to integrating the pixel values of $\vz$ that lie on each of the $M$ straight lines defined by vectors $\vt$ and $\vb$. If lenticules are detected correctly, then $z$ is equal to one on the boundary and zero everywhere else. Therefore, this metric is minimized when the fitted line lies exclusively on pixels located on a lenticule boundary. The objective function is, however, ill-posed as it does not guarantee that the entries in $\vt$ and $\vb$ fit different lenticules in the image. To address this problem, we need specific regularization terms.
Since we know that the width of the lenticules should be rather constant across the image, we can use this to enforce two regularization terms. We can penalize large deviations from the average lenticule width, as well as large width variations between consecutive lenticules. To this end, we formulate the following quadratic penalties:
\begin{equation}
     r_1(\vt,\vb) = (\mathbf{D}\vt - \hat{w} )^T (\mathbf{D} \vt  - \hat{w} ) + (\mathbf{D}\vb - \hat{w} )^T (\mathbf{D} \vb  - \hat{w} ) ,
     \label{eq:constr_1}
\end{equation}
\begin{equation}
     r_2(\vt,\vb) = \vt^T \mathbf{H}^T\mathbf{H} \vt + \vb^T \mathbf{H}^T\mathbf{H} \vb,
     \label{eq:constr_2}
\end{equation}
where $\mathbf{D}$ is a matrix of dimension $(M-1)\times M$ whose entries are defined as
\begin{equation}
     D(i,i) = -1, D(i,i+1) = 1
     \label{eq:D_def}
\end{equation}
and zero otherwise, whereas $\mathbf{H}$ is a matrix of dimension $(M-2)\times M$ whose entries are defined as:
\begin{equation}
     H(i,i) = 1, H(i,i+1) = -2, H(i,i+2) = 1,
     \label{eq:H_def}
\end{equation}
and zero otherwise, and $\hat{w}$ is a rough estimate of the lenticule width. Such estimate can simply be obtained with a Fourier analysis on $\vz$. The matrices $\mathbf{D}$, and $\mathbf{H}$ basically perform a first order, and second order respectively, finite difference on the vectors $\vt$ and $\vb$.
The full optimization can be written as:
\begin{equation}
     \underset{\vt,\vb}{\text{minimize}}\ \ \ - \sum_{m=0}^M \sum_{h=0}^H z(h,b_m + \frac{h t_m}{H}) + \lambda_1 r_1(\vt,\vb) + \lambda_2 r_2(\vt,\vb),
     \label{eq:lent_ext2}
\end{equation}
where $\lambda_1$ and $\lambda_2$  are positive parameters that balance the strength of the regularization terms. Note that the goal of the first regularization step is mainly to avoid fitting multiple vectorial lenticules on the same image location, whereas the second term enforces a certain smoothness on the width of nearby lenticules width.
The optimization problem of Eq.~\eqref{eq:lent_ext2} can effectively be solved with gradient based optimization methods. However,
because of the term $z(\cdot)$, the objective function is non-convex and has multiple local minima. As a result, it is important to provide a good initial guess to the gradient method in order to avoid being trapped in a local minimum. Fortunately, we observed that a simple peak detector applied to the output of the network to initialize $\vt$ and $\vb$ provides a rough estimate that is very effective, i.e., we did not observe any failure cases.

\begin{figure}[t]
    \centering
    \includegraphics[width=0.75\columnwidth]{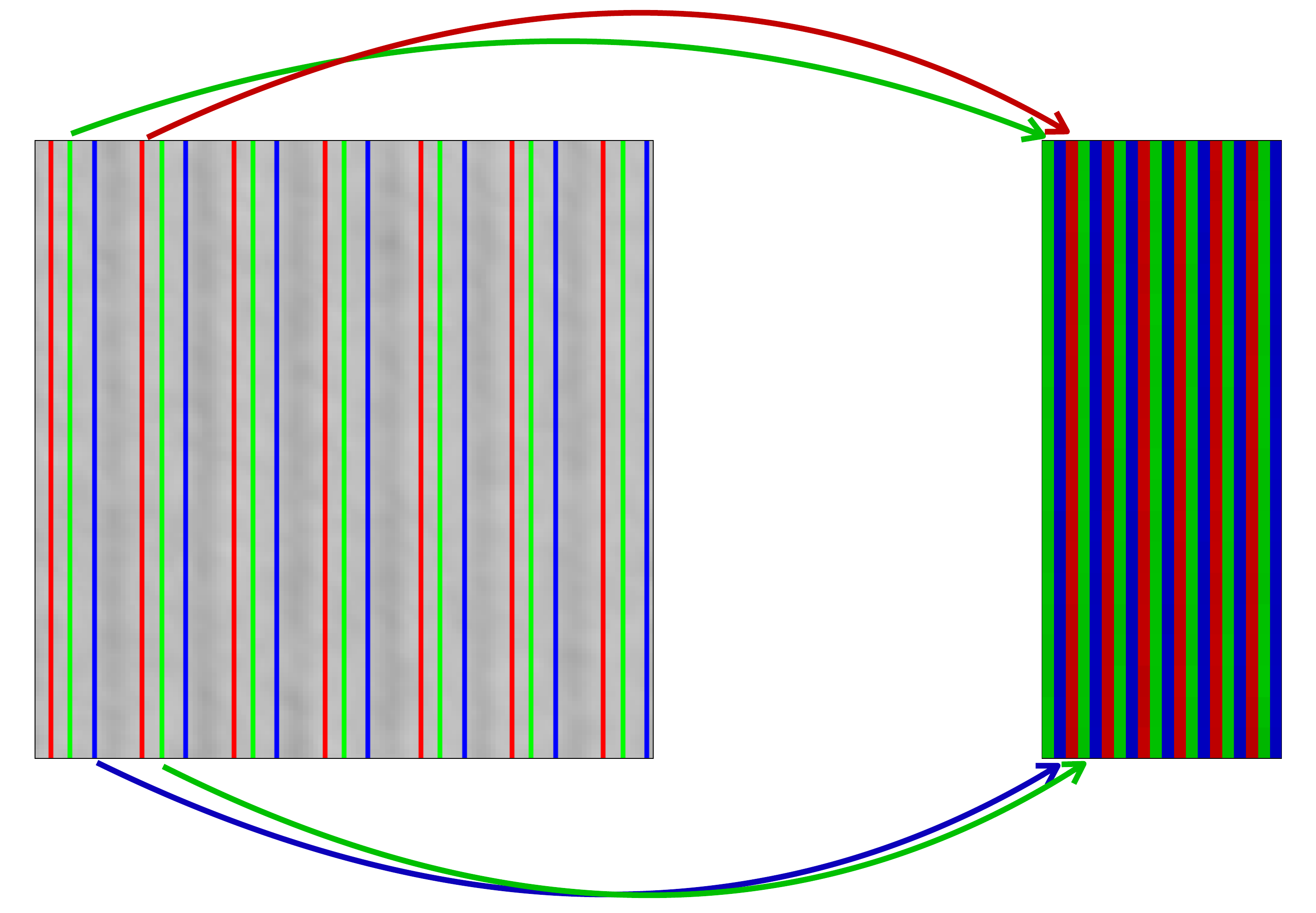}
    \caption{RGB value extraction. Left: detail of an input grayscale image with the RGB location within each lenticule highlighted. Right: The RGB columns are extracted and used to build the striped color image.}
    \label{fig:mosaic_build}
\end{figure}

The optimal values of $\vt$ and $\vb$ can be used to extract the RGB values from the input image. We simply need to iterate through all the entries of the vectors, which define the different lenticules, and extract the pixel values corresponding to the red, green and blue colors, see Fig.~\ref{fig:mosaic_build}. By carrying out this operation for all the rows of the input image $\vx$, of shape $H\times W$, we obtain a new image $H \times 3(M-1) \times 3$. Note that $M$ is the number of detected lenticule boundaries and we extract only lenticules that are entirely contained in the frame. The newly constructed image has three channels and while it has the same number of rows, the width is reduced because there are only three pixels per lenticule width. In order to rescale the image to the original proportion, we down-sample the image vertically to $H'=\frac{3 (M-1) H}{W}$. We then obtain an image of shape $H' \times 3(M-1) \times 3$ that has the same proportions of the input image and only one color channel is defined for each column.

\begin{figure}[t]
    \centering
    \includegraphics[width=\columnwidth]{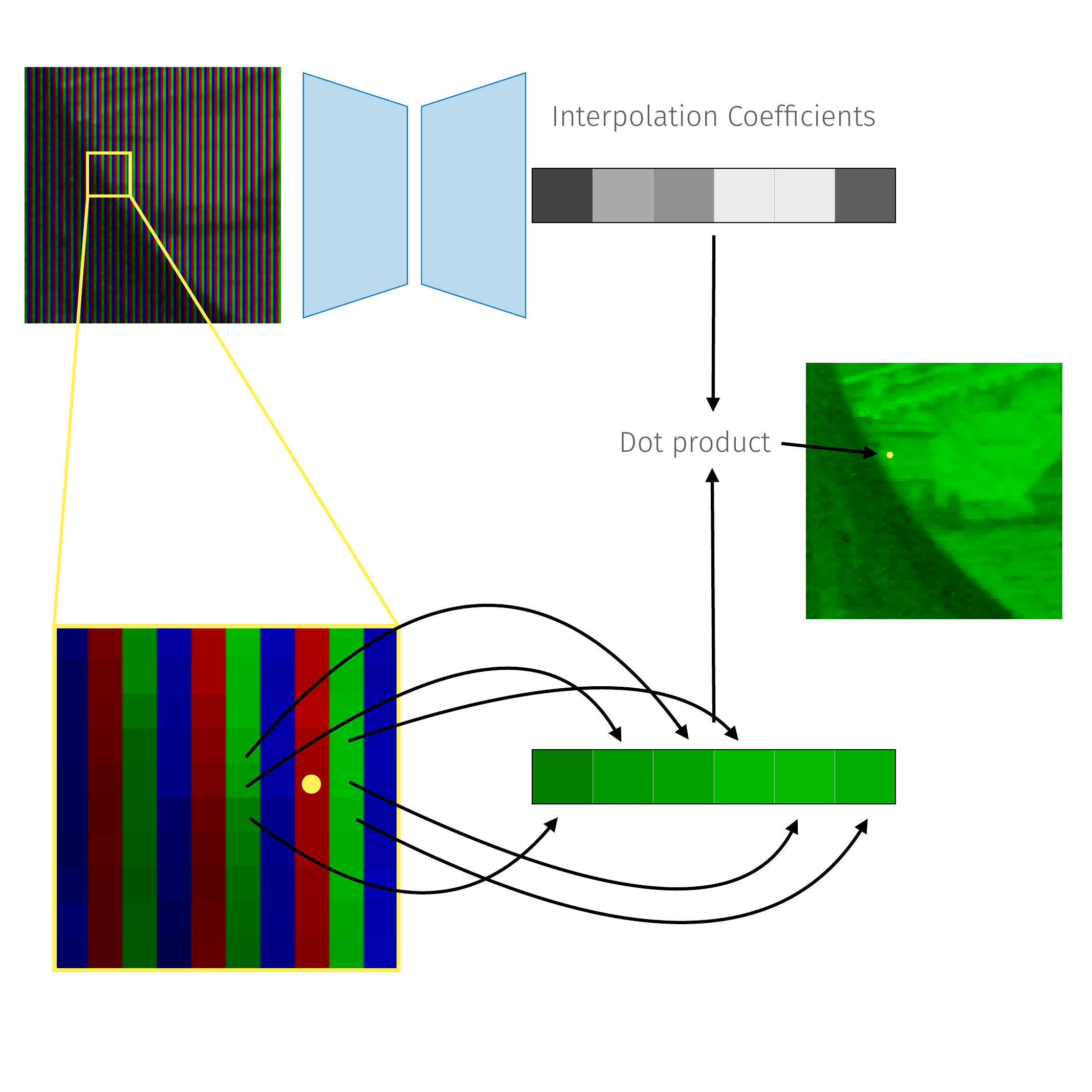}
    \caption{Detailed diagram of the Color reconstruction process. For each pixel, and for each color channel, the network outputs six positive weights that sum to one, these values represent the interpolation coefficients. For each pixel location in the input image the 6 closest values for each channel are extracted and concatenated in the feature dimension creating a feature map of shape $H'\times W' \times 6$ for each color channel. By performing a dot products in the feature dimension for each color channel we obtain the final RGB triplet.}
    \label{fig:color_rec}
\end{figure}

\subsection{Color Reconstruction}

One trivial way to fill the voids in the color channels is to use  some conventional interpolation technique, such as linear or cubic, between the valid pixels of each color channel. We instead decide to pursue an alternative approach, which proved to be effective in similar scenarios. It consists of training a CNN in order to fill the voids, similarly to what is done in recent super-resolution and demosaicing  methods~\cite{deep_demosaic_1,deep_demosaic_2}. 
Since we do not have a sufficient amount of analog reconstructed color images from lenticular films to train a neural network, we need to find an alternative training procedure. Fortunately, it is rather straightforward to create striped RGB images from fully colored ones. It is in fact sufficient to delete two RGB channels for each column of the image to replicate images that resemble the output of the RGB channel extraction module in our pipeline (see Fig.~\ref{fig:pipeline}). We can then use this synthetic dataset to train a network that reconstructs the fully colored image.
%
At test time we can use the trained network to colorize the lenticular images. It is important to perform a proper data augmentation during training though to be sure that a potential domain shift between training set and test set does not hurt model performance. To have a more robust network, we perturb images with additive noise, and by randomly changing the contrast and brightness of the training images.

A straightforward colorization network implementation would simply output a 3-channel image with the same spatial dimension of the input. In that case, however, the network would be able to colorize each pixel of the image with any possible color. Since we are still concerned about preserving the historical truthfulness of the colors, such a network architecture could potentially alter the hue of the picture. To alleviate this problem, we design a network that can colorize an output pixel by using exclusively a convex-combination of the closest valid pixels of the same color. We can think of this as the network is predicting the interpolation coefficients to reconstruct each pixel color (Fig.~\ref{fig:color_rec}). In our design, we decide to use the six closest neighbors of the same color for each output pixel\footnote{On a pixel column where one of the RGB values is available, we exclusively use the color values located on that column.}

In order to train the network we use a combination of $l_1$ loss and adversarial loss:
\begin{align}
     \underset{\vtheta}{\text{min.}}\ \underset{\varphi}{\text{max.}}\ \ \ &\gamma \mathbb{E}_{\vx,\vy_\text{gt}} [l_1(\vy (\vx),\vy_\text{gt})] + \nonumber\\ &\mathbb{E}_{\vx} [\log(1 - f_D(\vy (\vx)))] + 
     \nonumber\\ 
     &\mathbb{E}_{\vy_\text{gt}} [\log(f_D(\vy_\text{gt}))],
     \label{eq:col_loss}
\end{align}
where $\theta$ represent the learning parameters of the colorization network, $\varphi$ the parameters of the discriminator network, and $\gamma$ is a weight parameter that balances the two losses. Using an adversarial loss allows us to improve the quality of the reconstructed image as we are able to obtain sharper output images. 

\section{Experiments}\label{sec:exp}

\subsection{Datasets}
Our lenticular dataset consists of 2010 scanned frames of 12 lenticular films. Each frame is cropped to exclude the film perforations from the image. The average size of the images is about $2550 \times 3650$ pixels. Lenticule width can range from 12 to 20 pixels depending on the spatial resolution of the scan. This dataset is composed of images that were correctly colorized by doLCE~\cite{Reuteler2014}. Segmentation label masks that identify the lenticule boundaries are available alongside the grayscale input images. Note that such boundaries may not always perfectly correspond to the real ground truth locations, but nevertheless represent predictions that are sufficiently good to reconstruct the image colors. Because doLCE can only deal with perfectly vertical lenticules, we perform random rotations to train the network to generalize to tilted lenticules. We split the dataset into train, validation, and test subsets with 1810, 100, 100 images in each subset, respectively. During the training of the lenticule detection network, images are cut into patches of $256\times256$ pixels and a random rotation of maximum $1^\circ$ is applied to the input-output pair. No rotation is applied during test time and the input image is cut into patches, processed by the network, and then re-combined to form the whole image prediction. Moreover, we have access to additional images where the doLCE algorithm led to unsuccessful colorization as it failed to detect the correct lenticule boundaries. In these scans the lenticules were either tilted so that the vertical fit of doLCE performed poorly, or the images presented some challenging parts (e.g., extended dark areas) which would require the usage of some regularization technique. These additional images will be used to verify whether the proposed method is indeed more robust than the doLCE algorithm.

To train the colorization network we leverage a second dataset \emph{div2K}~\cite{div2K}. This dataset is composed of 1000 high resolution RGB images with a large diversity of contents. We first randomly crop the images with patches of $256\times256$ pixels and artificially create a stripe-like structure by deleting two of the RGB channels for each column to obtain the same pattern of the original lenticular images. Random additive noise, contrast variation, and brightness variation of the images are performed during training to make the colorization network more robust.
%
At test time, the model is used to colorize the lenticular images, for which no real groundtruth colorized version exists. The input images are cut into  patches and then recombined to obtain a full image prediction.

\subsection{Implementation Details}
The lenticule detection network has a U-Net structure~\cite{unet} with a ResNet encoder~\cite{resnet} pretrained on ImageNet~\cite{imagenet}. The model has been implemented using PyTorch~\cite{pytorch} and the Segmentation Models Package~\cite{smp}.
The architecture of the colorization network is the same as the one of the lenticule detection network. Only the final layer changes as in this case the output is not a binary segmentation map but the interpolation coefficient to reconstruct the color of each pixel. The two networks are trained in the same way: we use the Adam optimizer~\cite{adam} for 50 epochs with an initial learning rate of $0.001$ for the lenticule segmentation network and $0.005$ for the colorization network. The learning rate is decreased by half every 10 epochs for both networks. The $\gamma$ hyperparameter of Eq.~\eqref{eq:col_loss} is set to $10$. The rough lenticule width estimate $\hat{w}$ is obtained via a spectral analysis of the lenticule segmentation map in the horizontal direction. The spectrum shows a clear peak at the frequency corresponding to the periodical pattern of lenticular film.
The optimization problem of Eq.~\eqref{eq:lent_ext2} is solved using the L-BFGS-B method~\cite{lbfgsb} implemented in the scipy library~\cite{scipy}. The hyperparameters of the regularization terms are set to $\lambda_1=1.$ and $\lambda_2=10$. As we do not have access to actual lenticule locations groundtruth, it was not possible to perform any objective maximization for the selection of the hyperparameters. As a result the values were selected by a subjective analysis of the color reconstruction results, seeking for the values that led to lesser artifacts.
The output of this optimization is used to extract the RGB values inside each lenticule. We vertically interpolate the resulting image to maintain the same proportions like the input image as described in Section~\ref{sec:method}. We further apply a median filter in the vertical direction before the interpolation to reduce noise. 

\subsection{Lenticules Detection}

\begin{figure}
     \centering
     \begin{subfigure}[b]{0.2\textwidth}
         \centering
         \includegraphics[width=\textwidth]{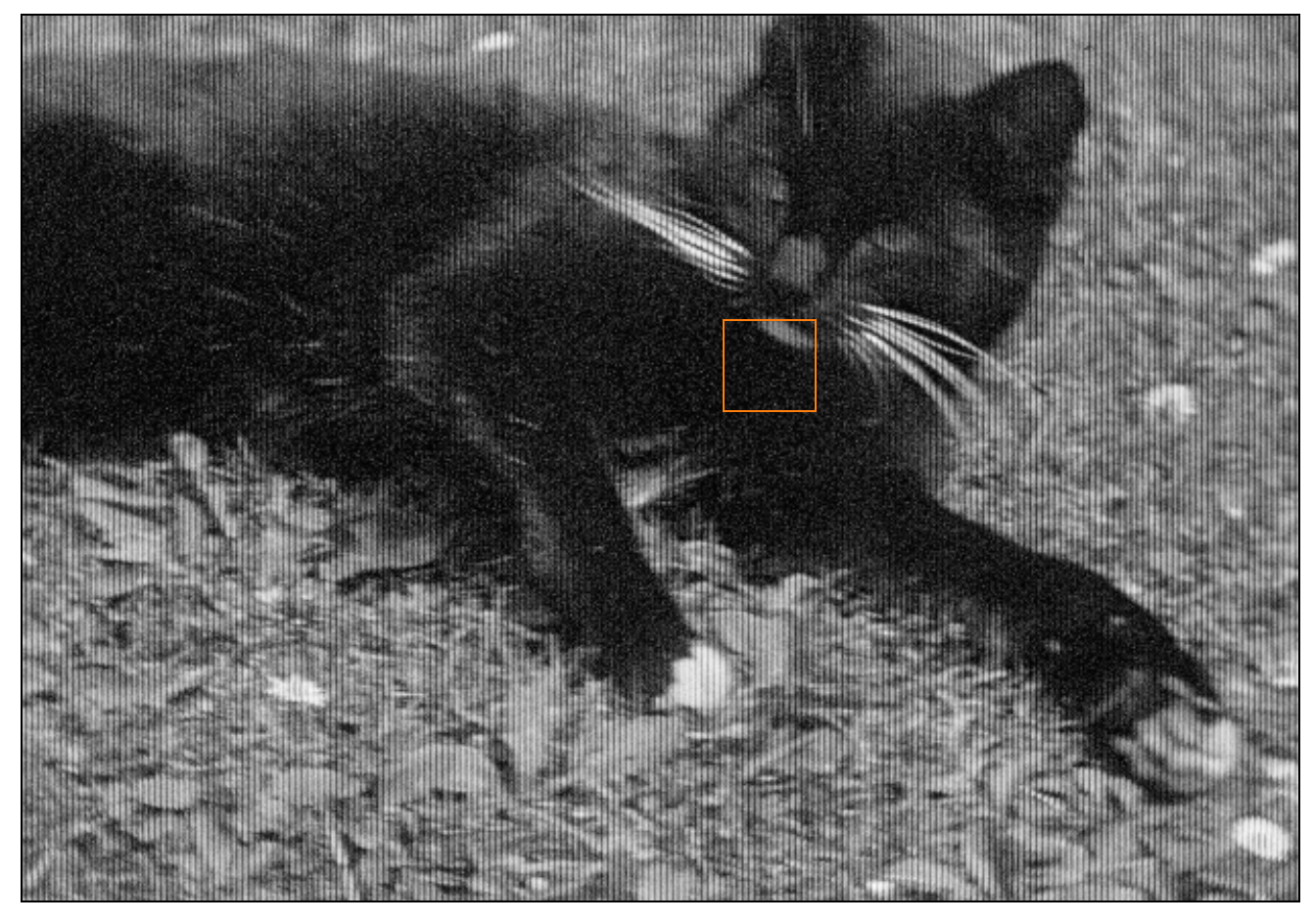}\\
         \vspace{.4cm}
         \caption{}
     \end{subfigure}
     \begin{subfigure}[b]{0.19\textwidth}
         \centering
         \includegraphics[width=0.9\textwidth]{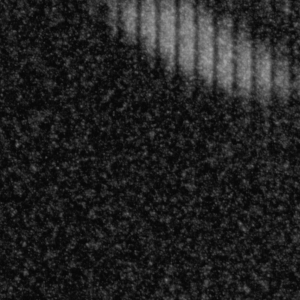}
         \caption{}
     \end{subfigure}\\
     \begin{subfigure}[b]{0.19\textwidth}
         \centering
         \includegraphics[width=0.9\textwidth]{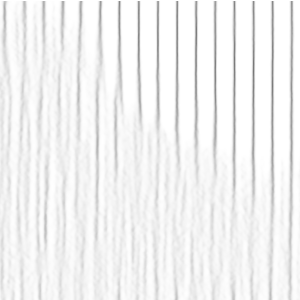}
         \caption{}
     \end{subfigure}
     \begin{subfigure}[b]{0.19\textwidth}
         \centering
         \includegraphics[width=0.9\textwidth]{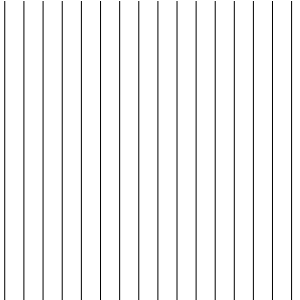}
         \caption{}
     \end{subfigure}
        \caption{(a) input image; (b) detail of the input image; (c) lenticule boundaries predicted by the network; (d) lenticule boundaries after the refinement module.}
        \label{fig:len_pred}
\end{figure}

In Fig.~\ref{fig:len_pred} we show an example of lenticule boundary detection and their refinement. The network can detect lenticule boundaries where they are visible and is robust to low contrast areas, too. In some cases, boundary detection becomes blurry (bottom left of Fig.~\ref{fig:len_pred}(c)) and too inaccurate for correct color reconstruction. Our refinement module successfully corrects for these effects by fitting straight lines (Fig.~\ref{fig:len_pred}(d)) vertically across the entire image. 

\begin{figure}
     \centering
     \begin{subfigure}[b]{0.24\textwidth}
         \centering
         \includegraphics[width=0.9\textwidth]{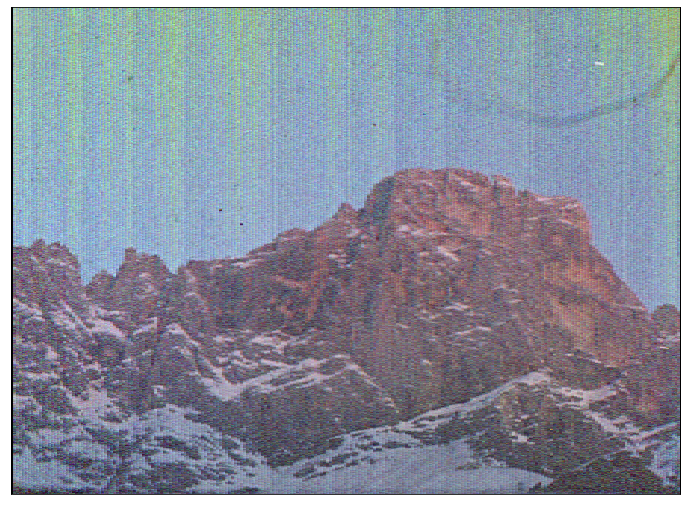}
     \end{subfigure}
     \begin{subfigure}[b]{0.24\textwidth}
         \centering
         \includegraphics[width=0.9\textwidth]{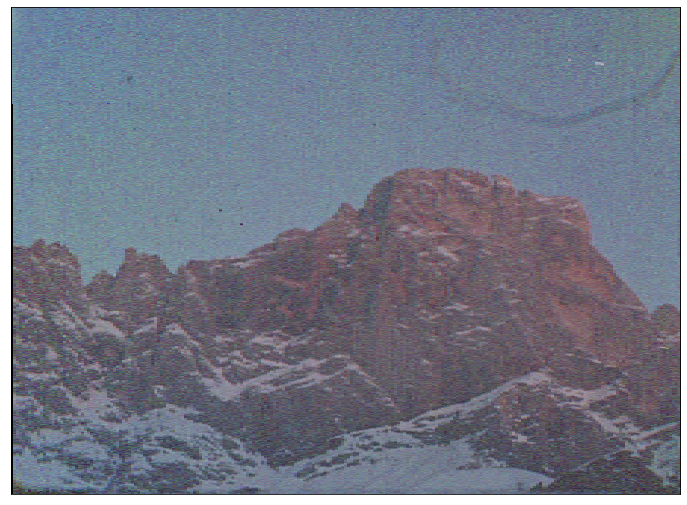}
     \end{subfigure}\\
     \begin{subfigure}[b]{0.24\textwidth}
         \centering
         \includegraphics[width=0.9\textwidth]{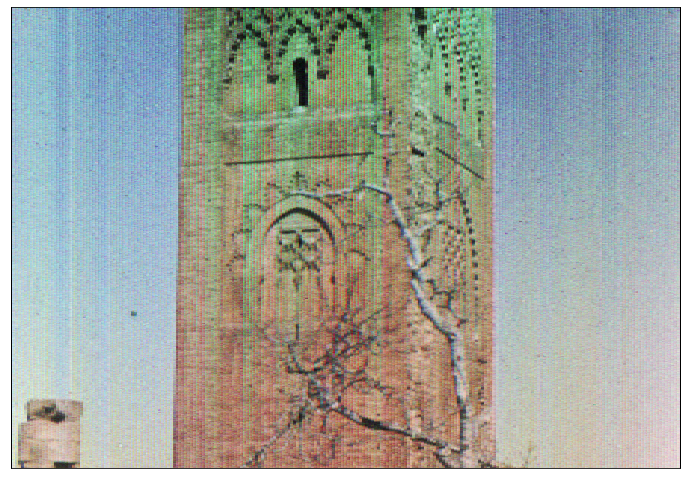}
     \end{subfigure}
     \begin{subfigure}[b]{0.24\textwidth}
         \centering
         \includegraphics[width=0.9\textwidth]{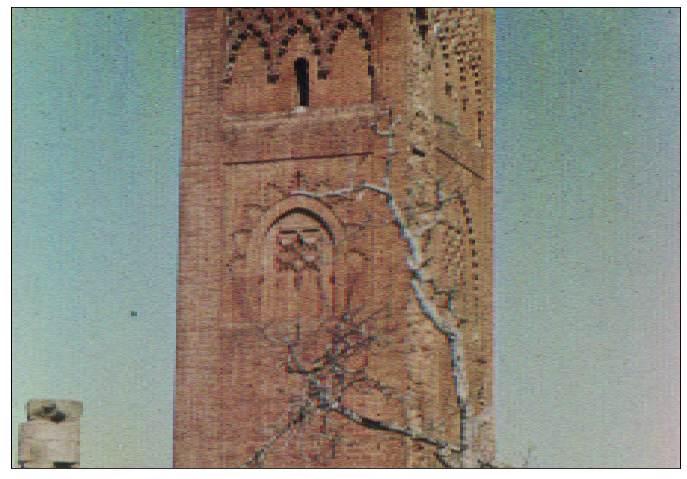}
     \end{subfigure}
          \begin{subfigure}[b]{0.24\textwidth}
         \centering
         \includegraphics[width=0.9\textwidth]{figures/failures/color_corr_rgb_RolleaAS_0221.jpg.png}
         \caption{}
     \end{subfigure}
     \begin{subfigure}[b]{0.24\textwidth}
         \centering
         \includegraphics[width=0.9\textwidth]{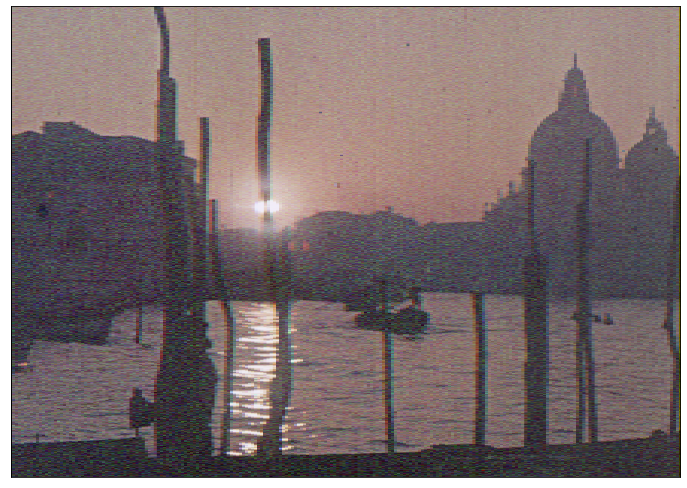}
         \caption{}
     \end{subfigure}
        \caption{(a) colorized images using the doLCE algorithm; (b) same images colorized using the lenticule detection pipeline proposed in this work. In this case we used the color reconstruction method of doLCE, in order to show that the color artifacts arise exclusively from the lenticules detection method.}
        \label{fig:dolce_failures}
\end{figure}

\begin{figure}[t]
    \centering
    \includegraphics[width=0.9\columnwidth]{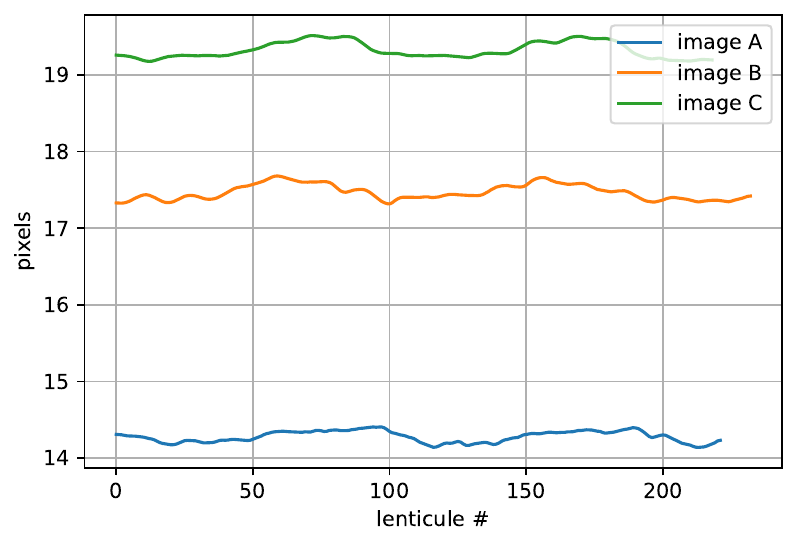}
    \caption{Horizontal variation of the average lenticule width for three images of the dataset.}
    \label{fig:lenticule_width}
\end{figure}

\begin{figure*}[h]
  \centering
  \begin{tabular}[c]{ccccc}
     \begin{subfigure}[c]{0.17\textwidth}
      \includegraphics[width=\textwidth]{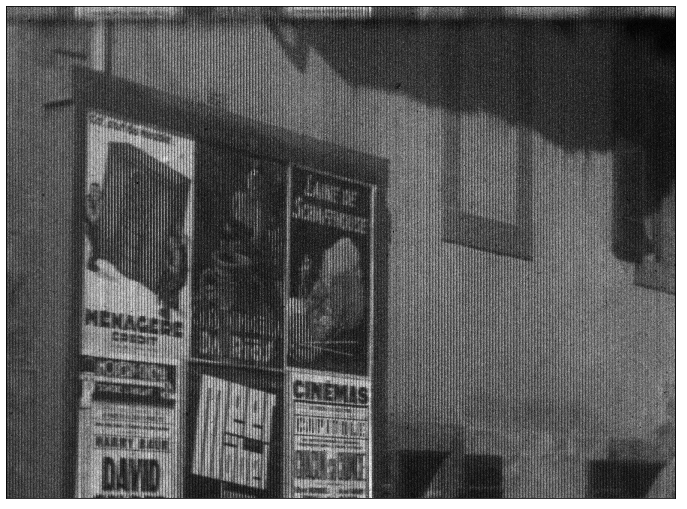}
    \end{subfigure}&
    \begin{subfigure}[c]{0.17\textwidth}
      \includegraphics[width=\textwidth]{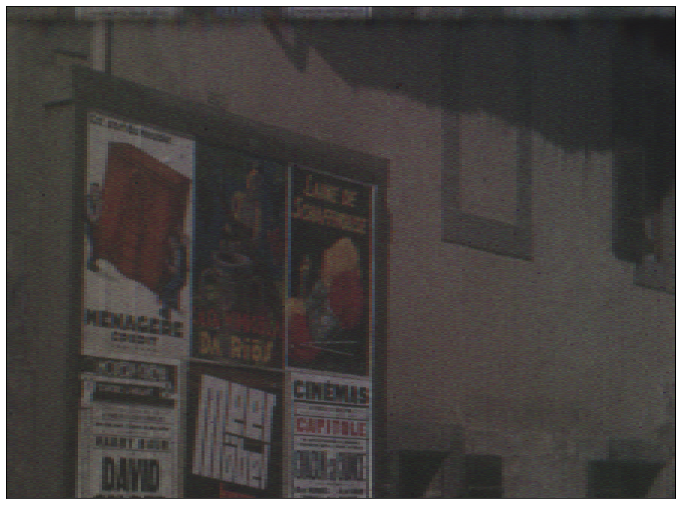}
    \end{subfigure}&
    \begin{subfigure}[c]{0.17\textwidth}
      \includegraphics[width=\textwidth]{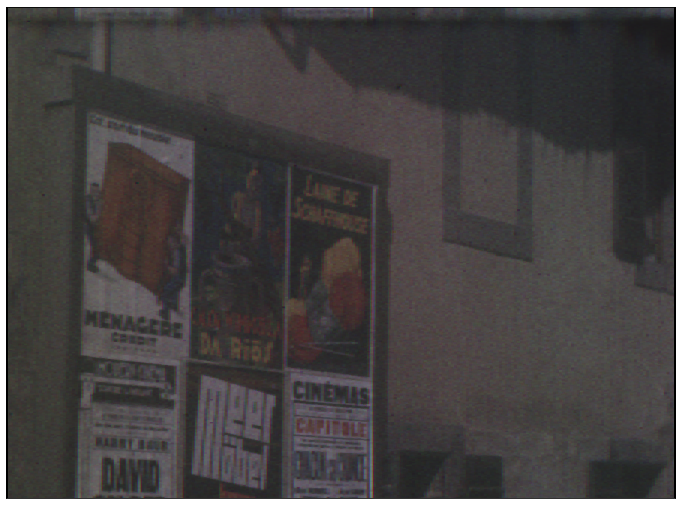}
    \end{subfigure}&
    \begin{subfigure}[c]{0.17\textwidth}
      \includegraphics[width=\textwidth]{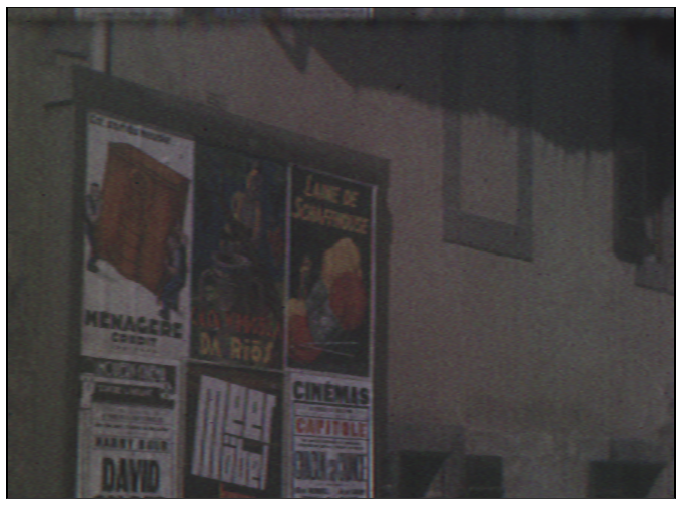}
    \end{subfigure}&
    \begin{subfigure}[c]{0.17\textwidth}
      \includegraphics[width=\textwidth]{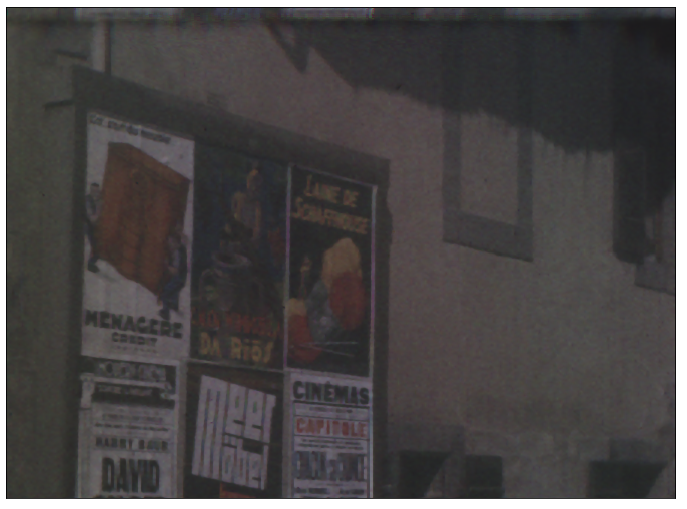}
    \end{subfigure}\\
    \begin{subfigure}[c]{0.17\textwidth}
      \includegraphics[width=\textwidth]{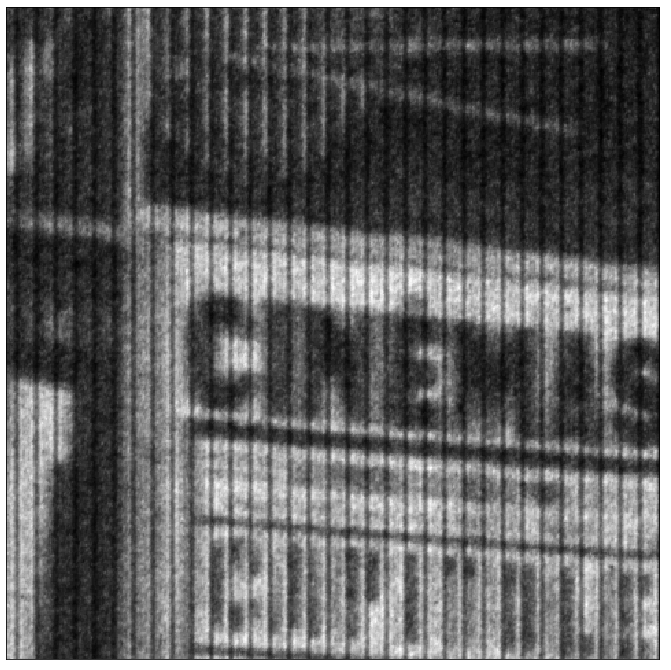}
    \end{subfigure}&
    \begin{subfigure}[c]{0.17\textwidth}
      \includegraphics[width=\textwidth]{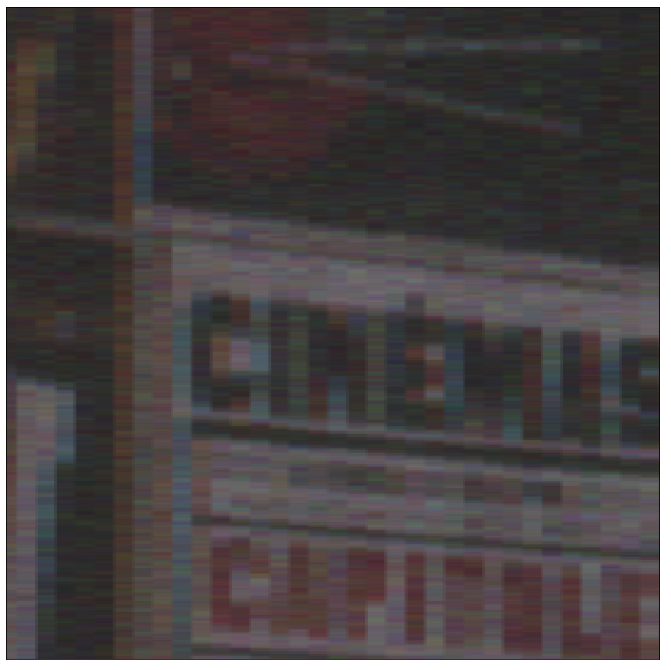}
    \end{subfigure}&
    \begin{subfigure}[c]{0.17\textwidth}
      \includegraphics[width=\textwidth]{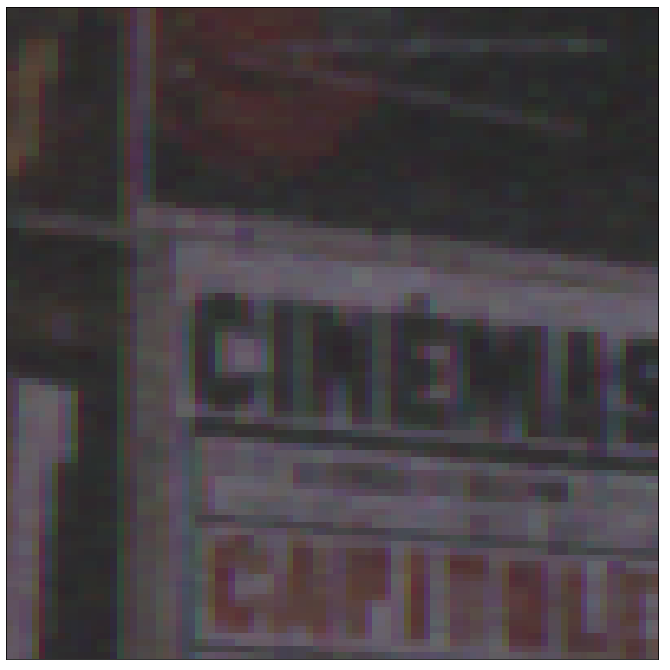}
    \end{subfigure}&
    \begin{subfigure}[c]{0.17\textwidth}
      \includegraphics[width=\textwidth]{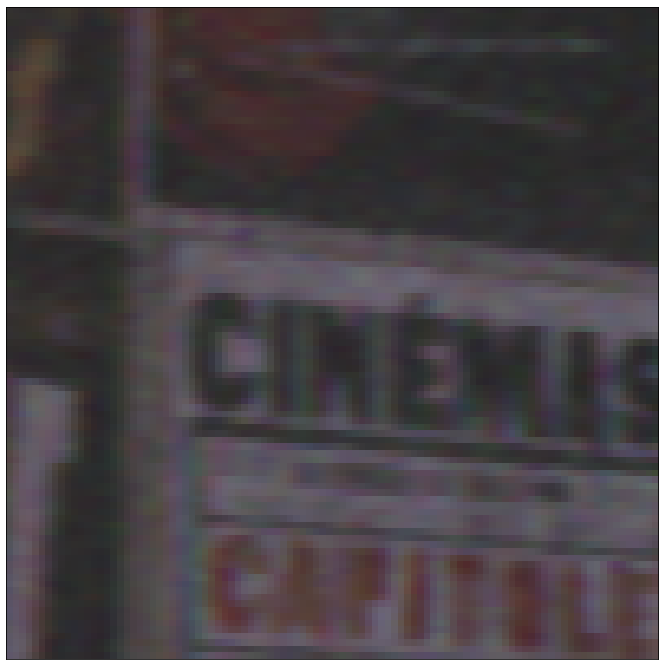}
    \end{subfigure}&
    \begin{subfigure}[c]{0.17\textwidth}
      \includegraphics[width=\textwidth]{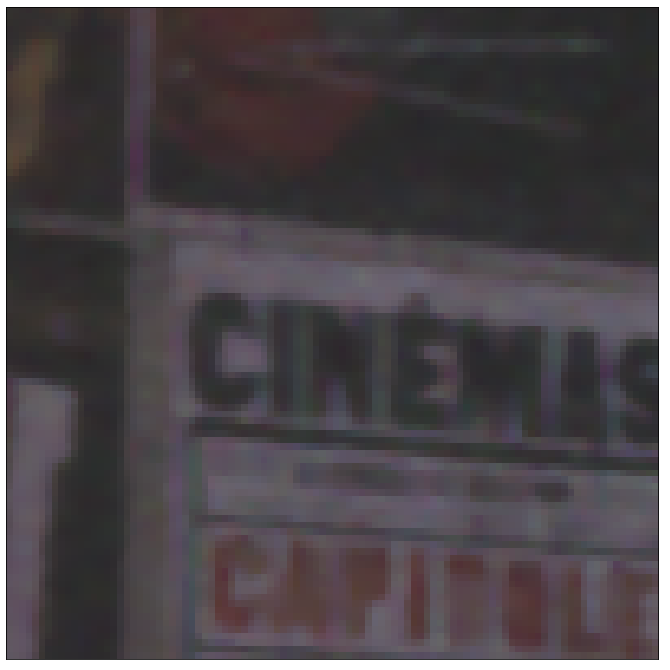}
    \end{subfigure}\\
   \begin{subfigure}[c]{0.17\textwidth}
      \includegraphics[width=\textwidth]{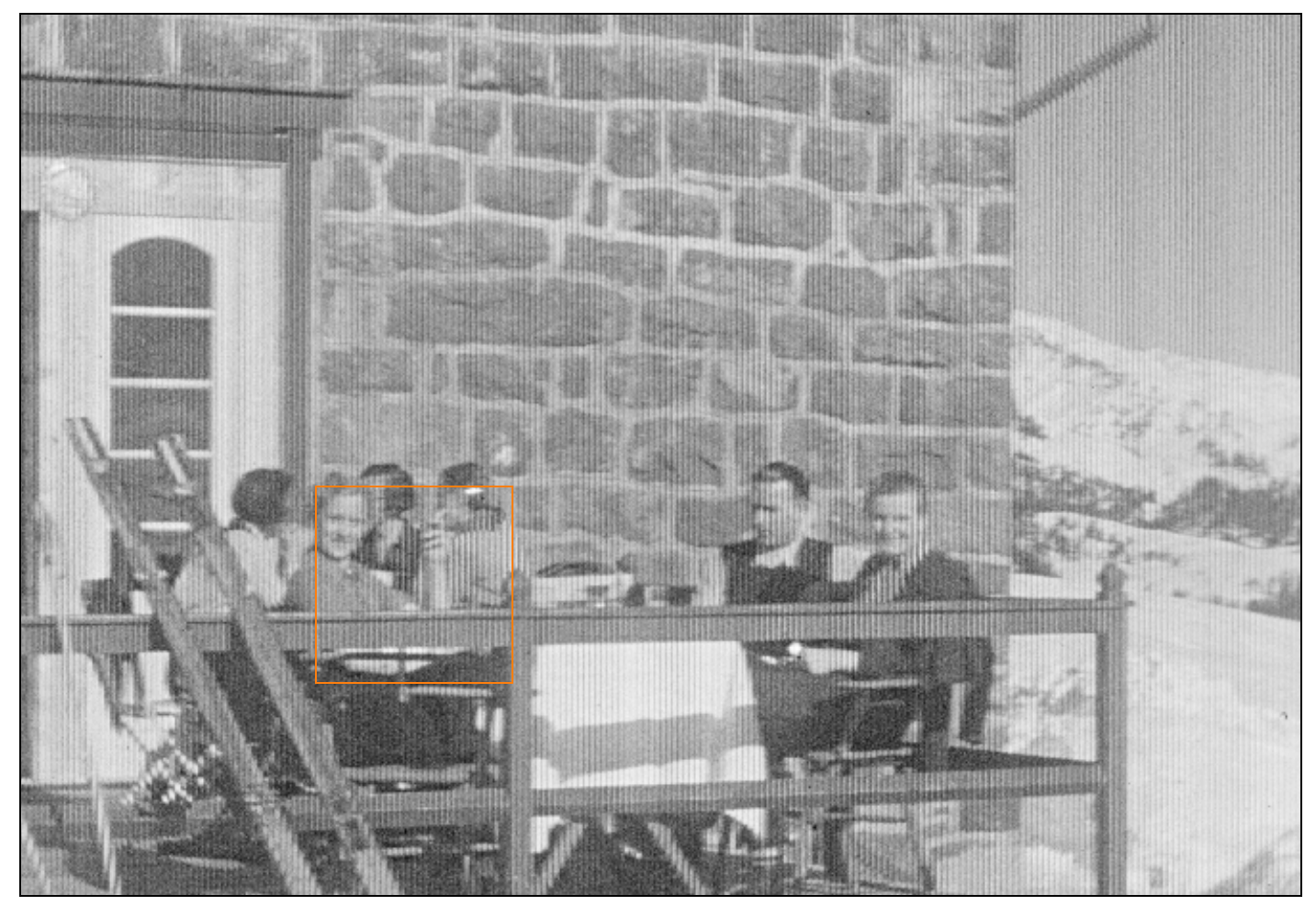}
    \end{subfigure}&
    \begin{subfigure}[c]{0.17\textwidth}
      \includegraphics[width=\textwidth]{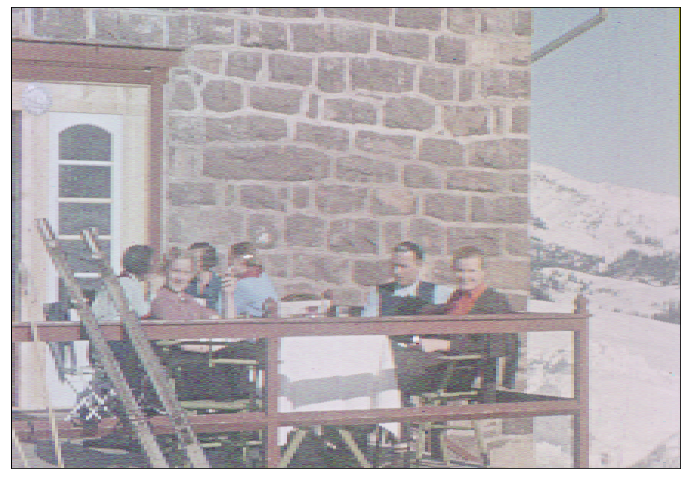}
    \end{subfigure}&
    \begin{subfigure}[c]{0.17\textwidth}
      \includegraphics[width=\textwidth]{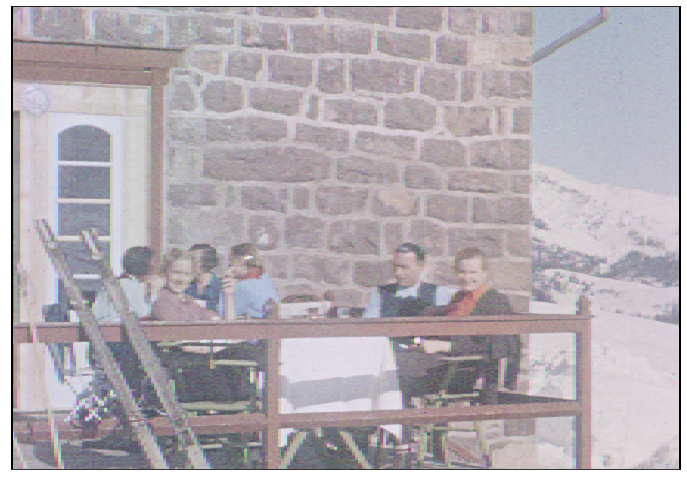}
    \end{subfigure}&
    \begin{subfigure}[c]{0.17\textwidth}
      \includegraphics[width=\textwidth]{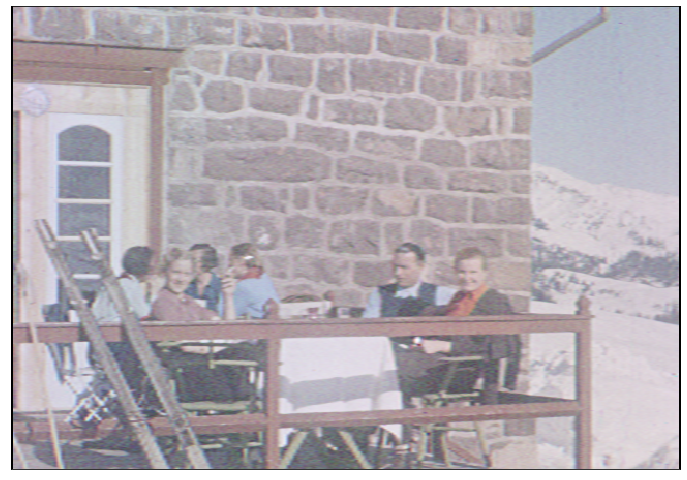}
    \end{subfigure}&
    \begin{subfigure}[c]{0.17\textwidth}
      \includegraphics[width=\textwidth]{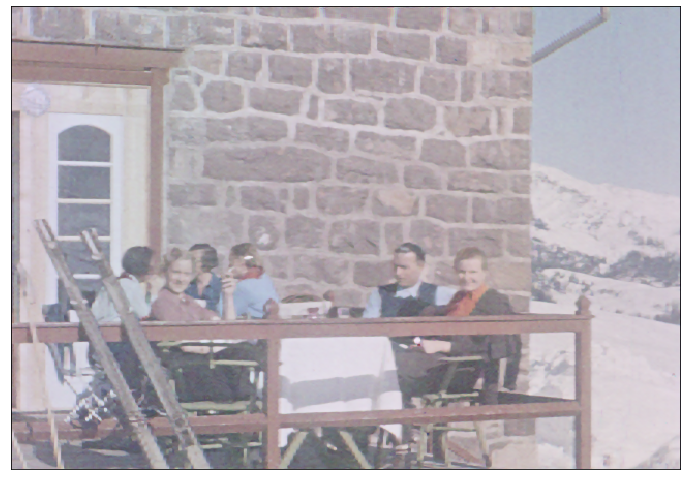}
    \end{subfigure}\\
    \begin{subfigure}[c]{0.17\textwidth}
      \includegraphics[width=\textwidth]{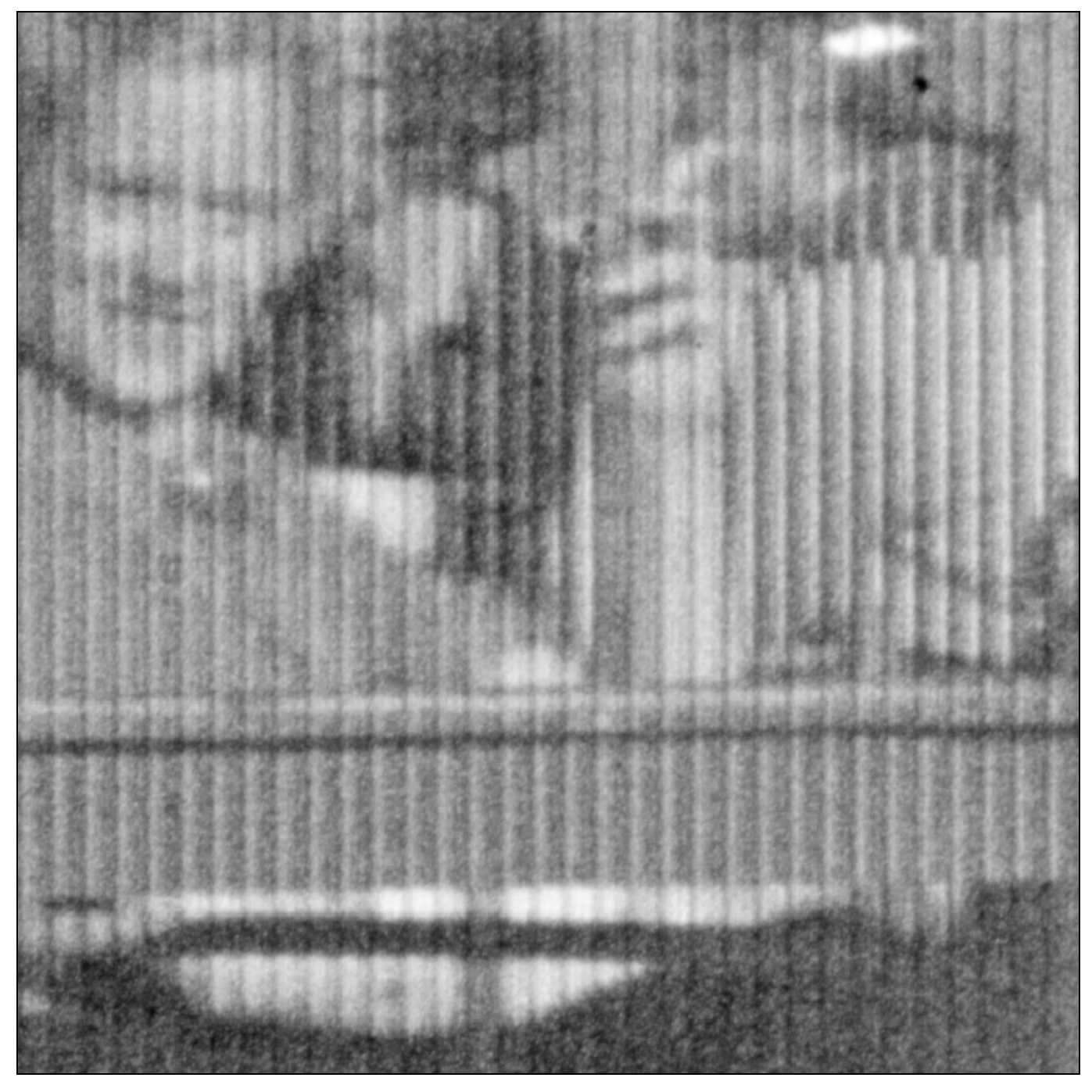}
      \caption{}
      \label{fig:color1_input}
    \end{subfigure}&
    \begin{subfigure}[c]{0.17\textwidth}
      \includegraphics[width=\textwidth]{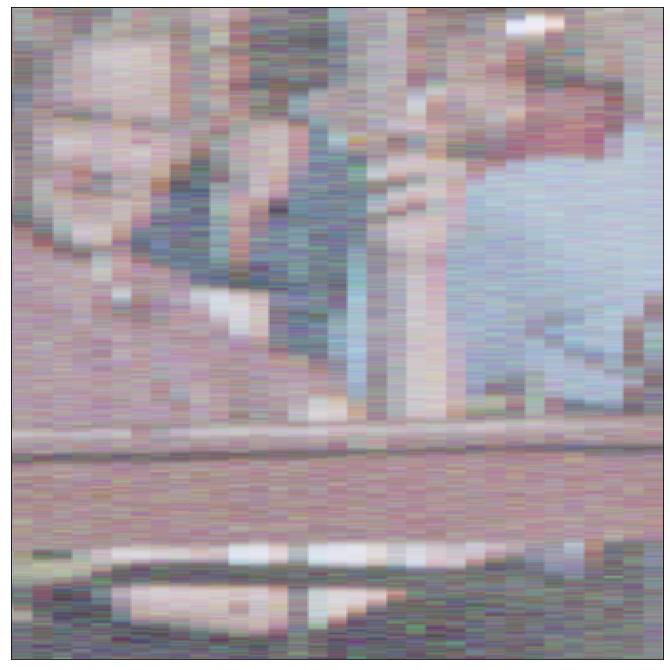}
      \caption{}
      \label{fig:color1_dolce}
    \end{subfigure}&
    \begin{subfigure}[c]{0.17\textwidth}
      \includegraphics[width=\textwidth]{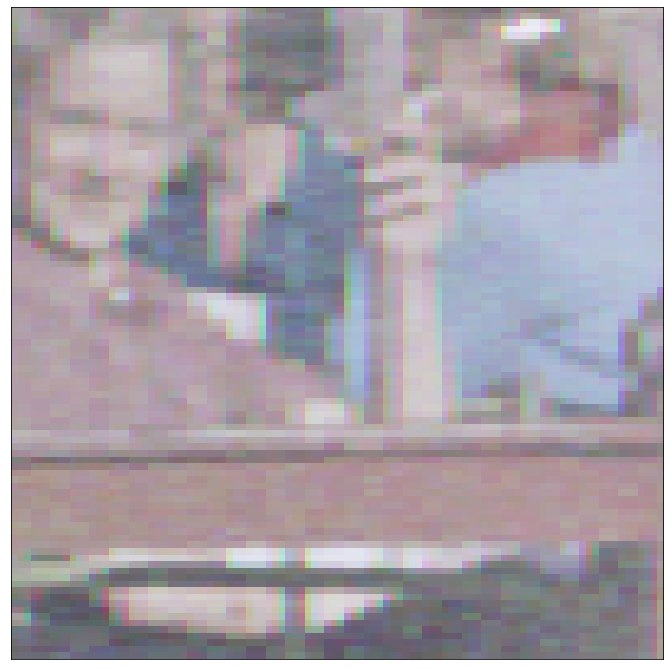}
      \caption{}
      \label{fig:color1_zero}
    \end{subfigure}&
    \begin{subfigure}[c]{0.17\textwidth}
      \includegraphics[width=\textwidth]{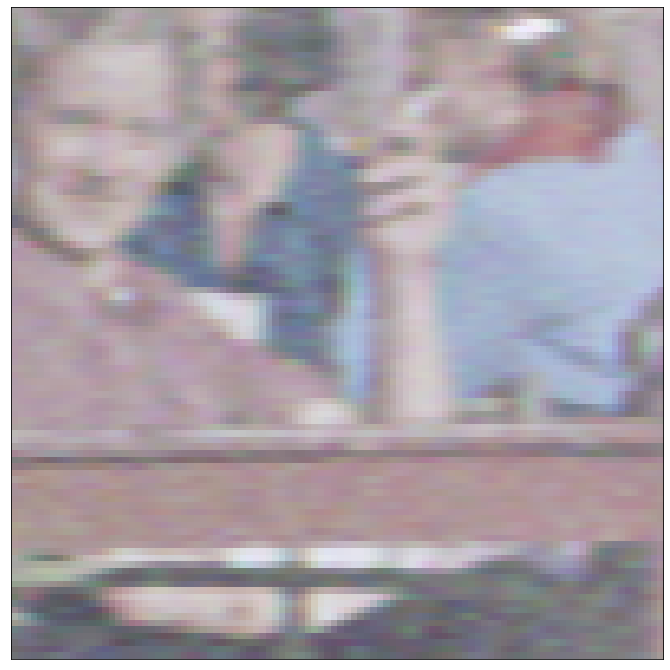}
      \caption{}
      \label{fig:color1_cubic}
    \end{subfigure}&
    \begin{subfigure}[c]{0.17\textwidth}
      \includegraphics[width=\textwidth]{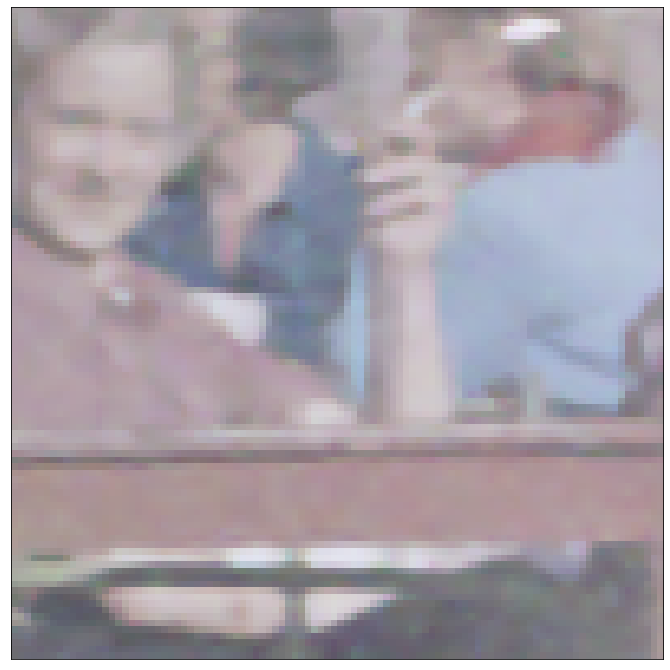}
      \caption{}
      \label{fig:color1_ours}
    \end{subfigure}\\
  \end{tabular}    
  \caption{Comparison of different color reconstruction techniques: (a) input image, (b) DoLCE algorithms, (c) nearest interpolation, (d) bicubic interpolation, (e) proposed methods.}
  \label{fig:color1}
\end{figure*}

Although we use the lenticule boundary predictions of doLCE to train the segmentation network, it is not entirely meaningful to use those in order to evaluate our method quantitatively. Recall that our goal is not to build a network that could replicate the old method but rather to create a network that is more robust, and able to deal with pictures where the old method failed to reconstruct the colors. 
The best way to validate successful lenticule detection of our approach is to look at images where doLCE failed. 
We show three of such examples in Fig.~\ref{fig:dolce_failures}, where our proposed method is able to correctly predict the lenticule boundaries even in images where the doLCE algorithm failed. 
In this case the failures can be associated to different causes. For the first two images (top and center) the artifacts are caused by the tilted lenticules. When lenticules that are slightly tilted are fitted with vertical ones we usually observe a color shift (green in this case) from the bottom to the top of the image. This is caused by the fact that if the detection is correct at the bottom (top) of the image it gets progressively misaligned when moving up (down), causing a color shift.
The artifact in the bottom image is instead of a different nature. We speculate that the high brightness of the sun behind the pole causes some confusion in the doLCE lenticule detection, whereas for the proposed model, the CNN is able to learn from the successful predictions how the lenticule grid usually looks like and detects the lenticule boundaries correctly.

Finally, we analyze how the lenticule width changes within one image along the horizontal direction. In Fig.~\ref{fig:lenticule_width} we show the evolution of the lenticule width for some images of the dataset. As it can be seen, the lenticules width is not constant and reveals a clear low frequency component where in some parts of the image lenticules are consistently larger (or smaller) then the average. We observed similar patterns for several images of the dataset. Because of these low frequency variations we cannot have a fixed lenticule width for the entire image as by doing so, a progressive misalignment would build up, leading to severe colorization artifacts.

\subsection{Color Reconstruction}

To evaluate the digital colorization process, we compare our proposed method (Fig.~\ref{fig:color1}(e)) with three other methods: the doLCE colorization process (Fig.~\ref{fig:color1}(b)), nearest interpolation (Fig.~\ref{fig:color1}(c)), and cubic interpolation (Fig.~\ref{fig:color1}(d)). Note that doLCE spreads the same RGB value over the entire lenticule width creating a visible stripy patterns where pixels throughout the horizontal direction of a lenticule have the same color value (Fig.~\ref{fig:color1}(b)). Nearest interpolation slightly improves the quality of color reconstruction at the cost of losing details (Fig.~\ref{fig:color1}(c)) that appear in the original grayscale image versions (Fig.~\ref{fig:color1}(a)). Our proposed learning-based method (Fig.~\ref{fig:color1}(e)) is able to preserve most details and provides overall sharper images compared to bicubic interpolation (Fig.~\ref{fig:color1}(d)).

\begin{figure}
    \centering
    \includegraphics[width=\columnwidth]{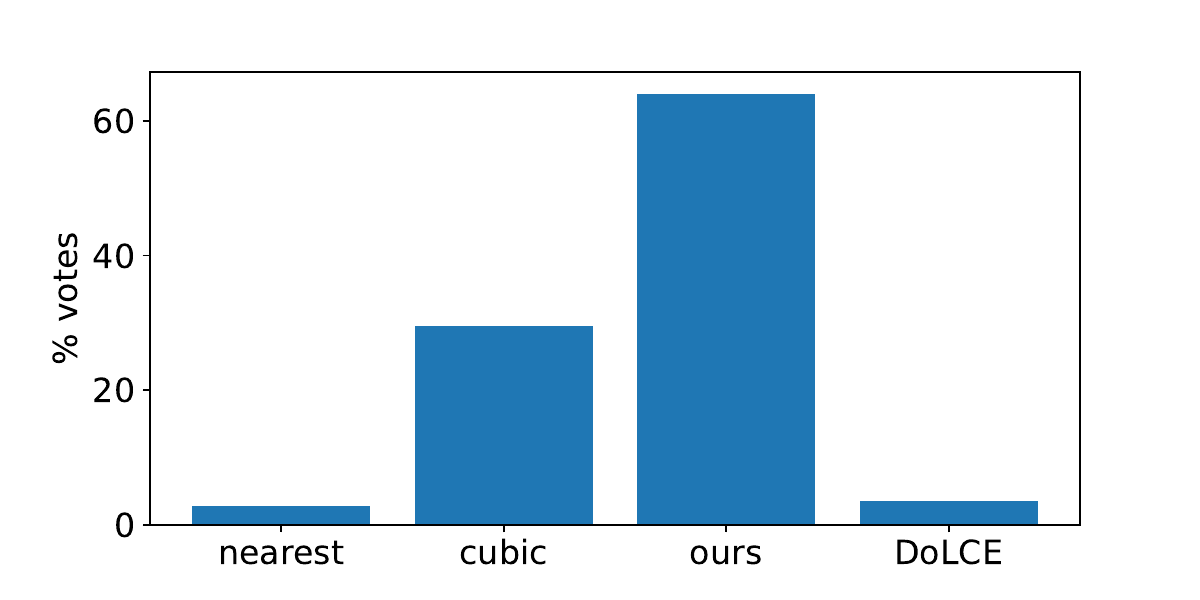}
    \caption{Results of the user study. The bar plot shows the percentage of the times each reconstruction method was selected as the best one by the participants.}
    \label{fig:user_study}
\end{figure}

Unfortunately it is not possible to have an objective evaluation of the different methods since we do not have access to a groundtruth reference images. We can however design a user study where we ask participants to evaluate which method in their opinion leads to the best looking image. To do this we selected $108$ images, from the test set plus some failures cases, we then displayed 4 color reconstructed images: the former doLCE colorization, the proposed lenticule detection network with cubic interpolation, the proposed lenticule detection network with nearest interpolation, and the full proposed method with the colorization network. For each single frame we asked the users to select, in their opinion, which color reconstruction looked better, according to level of details, sharpness, and absence of color artifacts. In total  1232 frames were rated by 33 participants. The results of the user study are shown in Fig~\ref{fig:user_study}. As it can be seen users selected about 60\% of the time the proposed method as the one generating the best looking picture. As expected the cubic interpolation ranked second, as it was selected as the best one about 30\% of times, followed by the doLCE method and nearest interpolation. 

\begin{figure}[t]
    \centering
    \includegraphics[width=0.48\columnwidth]{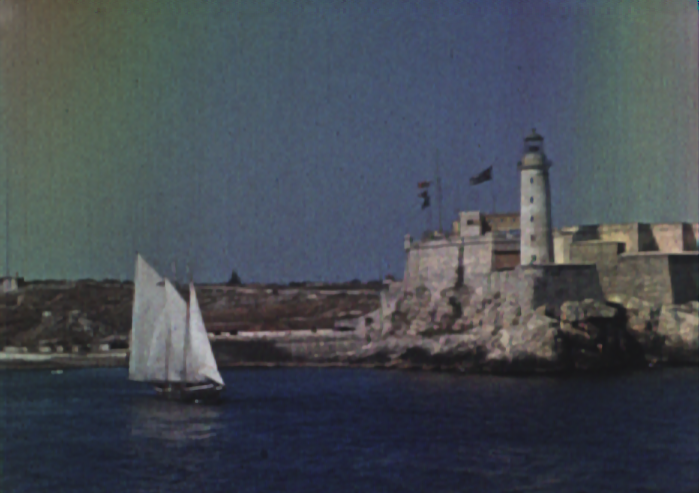}
    \includegraphics[width=0.48\columnwidth]{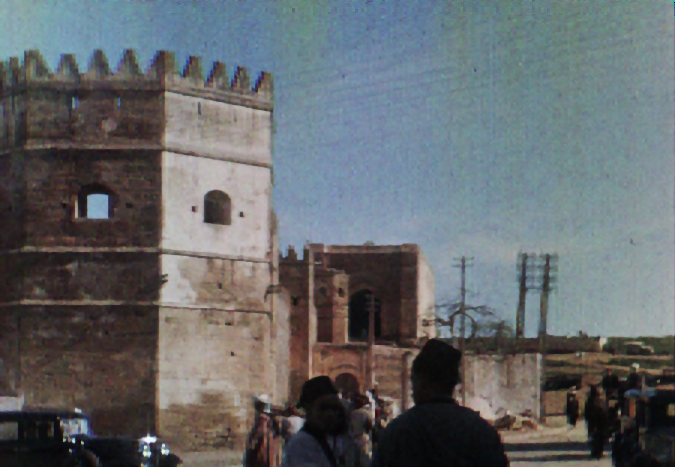}
    \caption{Failure cases for the color reconstruction process.}
    \label{fig:final_failure_cases}
\end{figure}

In Fig.~\ref{fig:final_failure_cases} we show some failure cases for the colorized images. In these images some color shifting is visible mainly the sky regions. This is caused by a misalignment between the actual lenticules and the predicted ones. In some cases large blue areas still pose a challenge for the overall method as the red color magnitude is much smaller than the blue one, and this makes the lenticule boundary look like it would be located slightly on the right of its actual position, leading ultimately to an erroneous color extraction. Unfortunately without highly accurate lenticule location groundtruth to use for training, overcoming this issue is quite difficulty; Fortunately we observed that failure cases are rare and lead to lesser artifacts than the doLCE method.

\begin{figure*}[h]
  \centering
  \begin{tabular}[c]{ccc}
     \begin{subfigure}[c]{0.31\textwidth}
      \includegraphics[height=2.3cm]{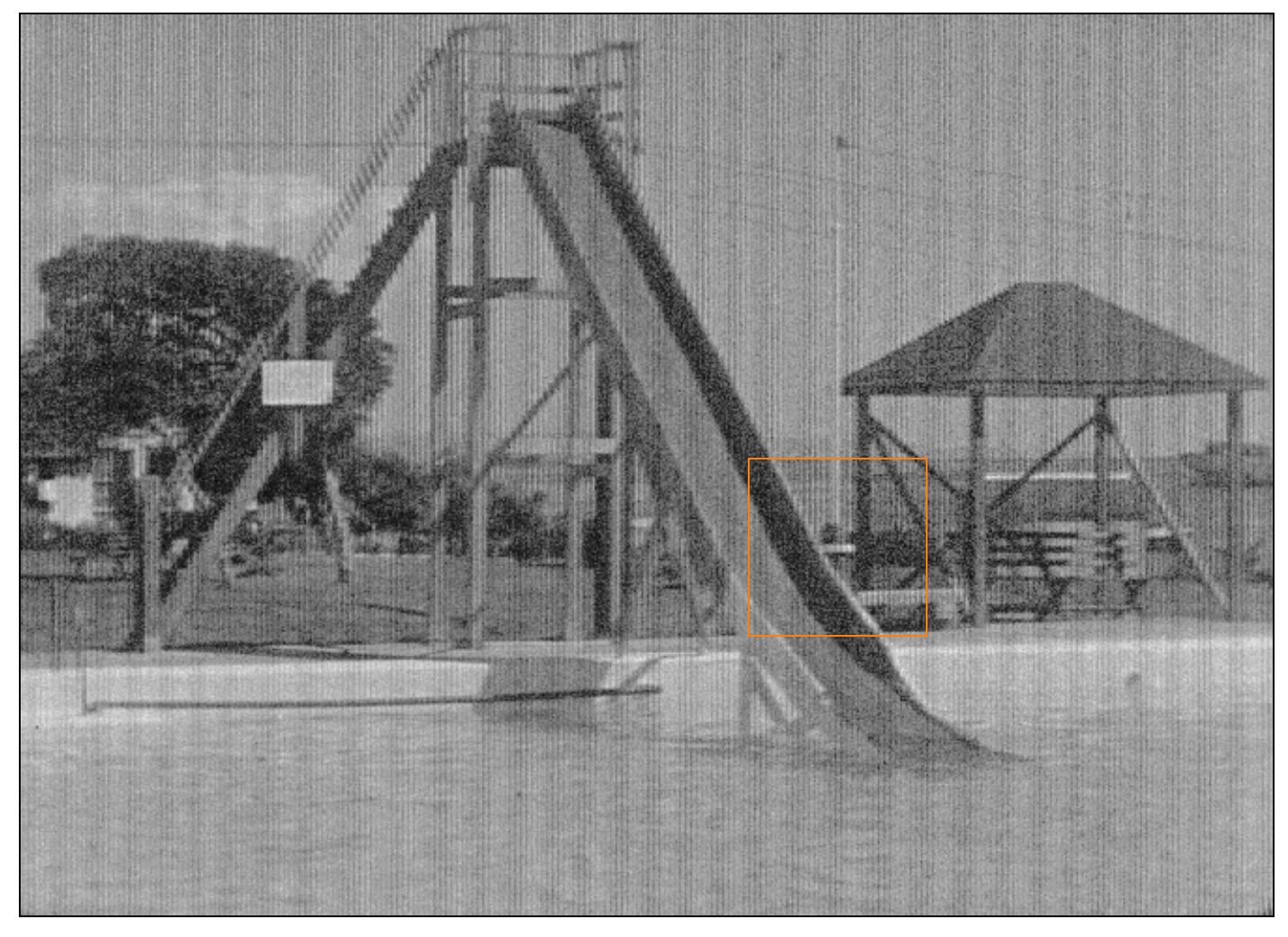}
    \hfill
      \includegraphics[height=2.3cm]{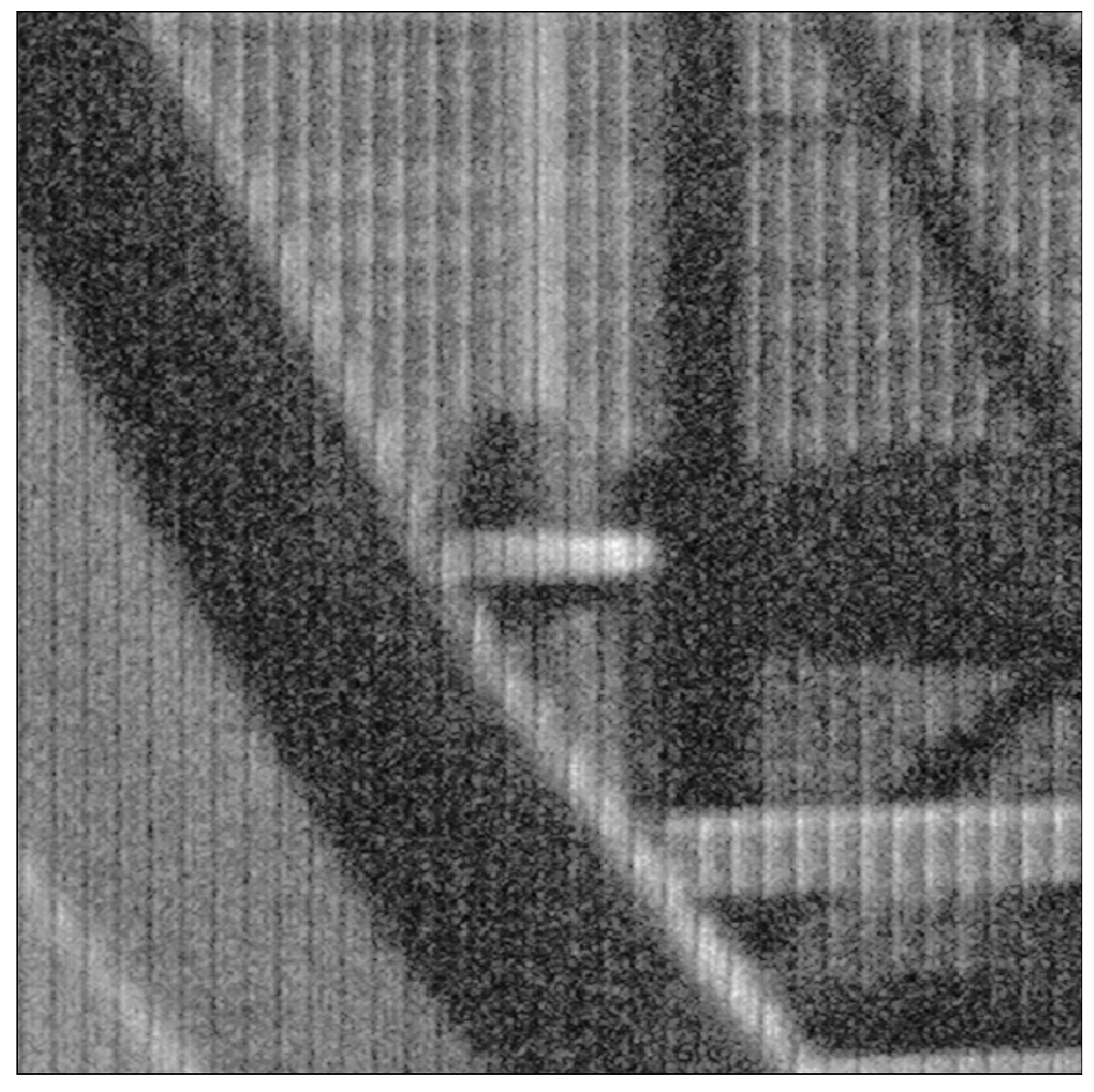}
    \end{subfigure}&
    \begin{subfigure}[c]{0.31\textwidth}
      \includegraphics[height=2.25cm]{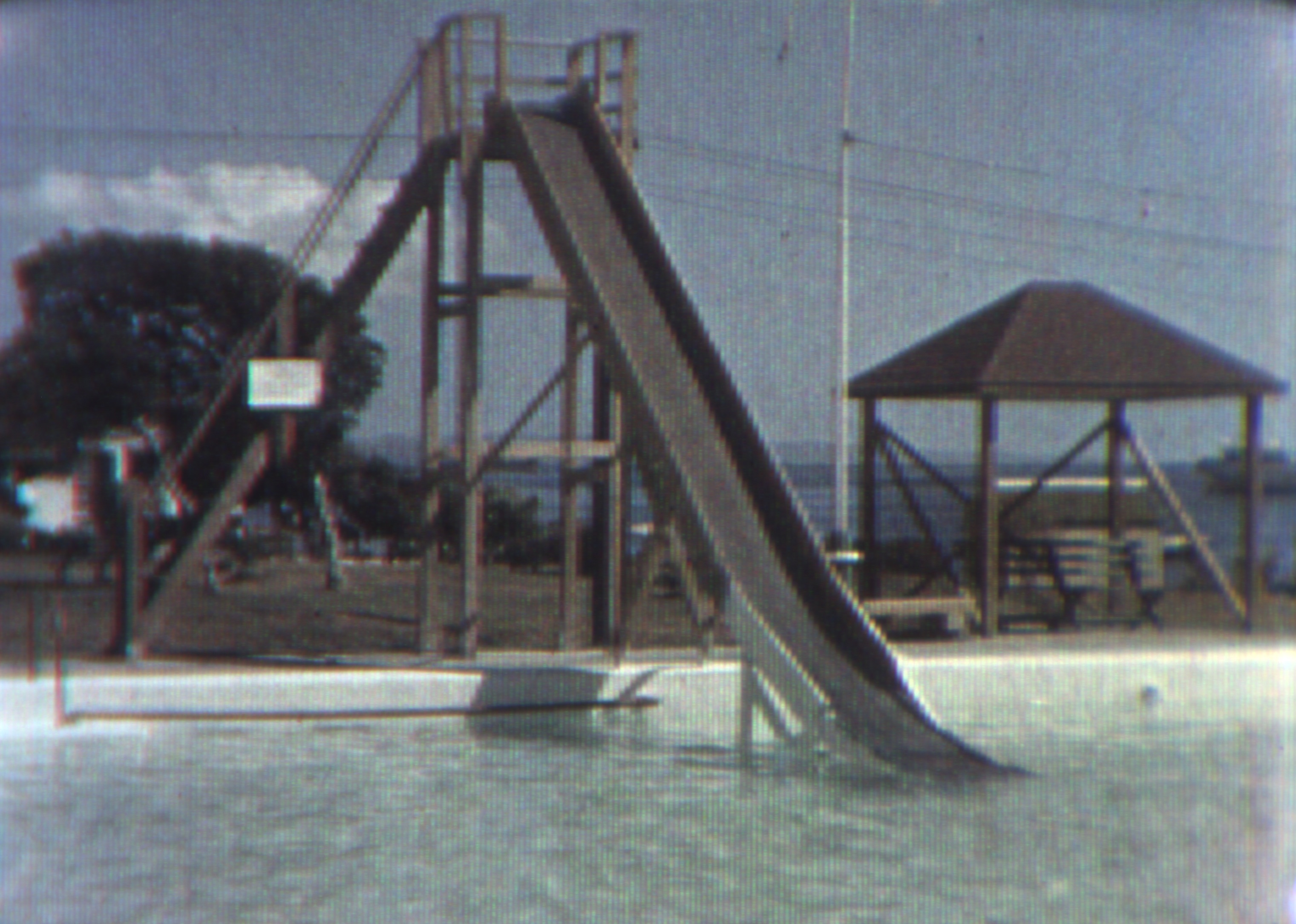}
          \hfill
\includegraphics[height=2.25cm]{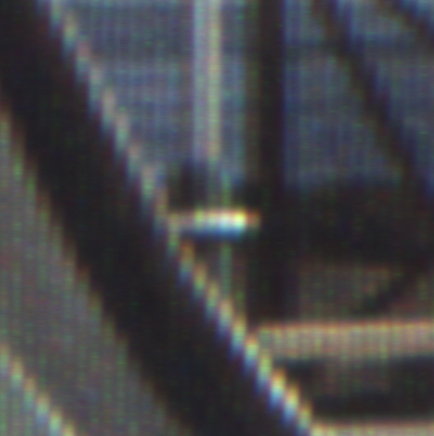}
    \end{subfigure}&
    \begin{subfigure}[c]{0.31\textwidth}
      \includegraphics[height=2.3cm]{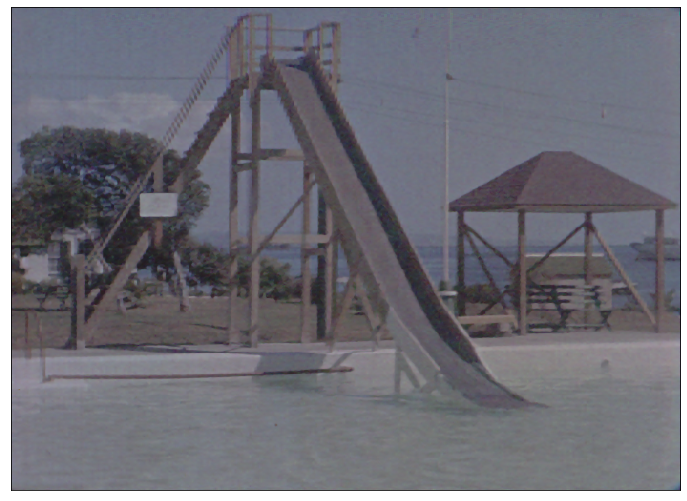}
          \hfill
\includegraphics[height=2.3cm]{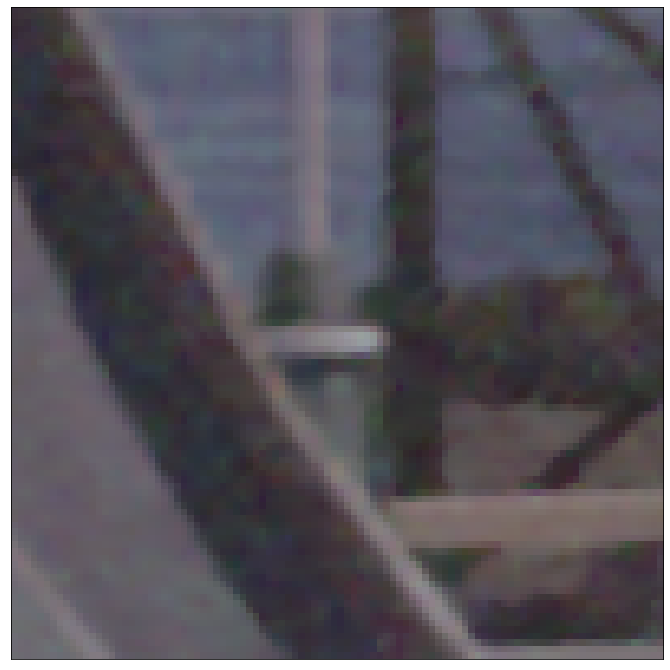}
    \end{subfigure}\\
    
    \begin{subfigure}[c]{0.31\textwidth}
      \includegraphics[height=2.3cm]{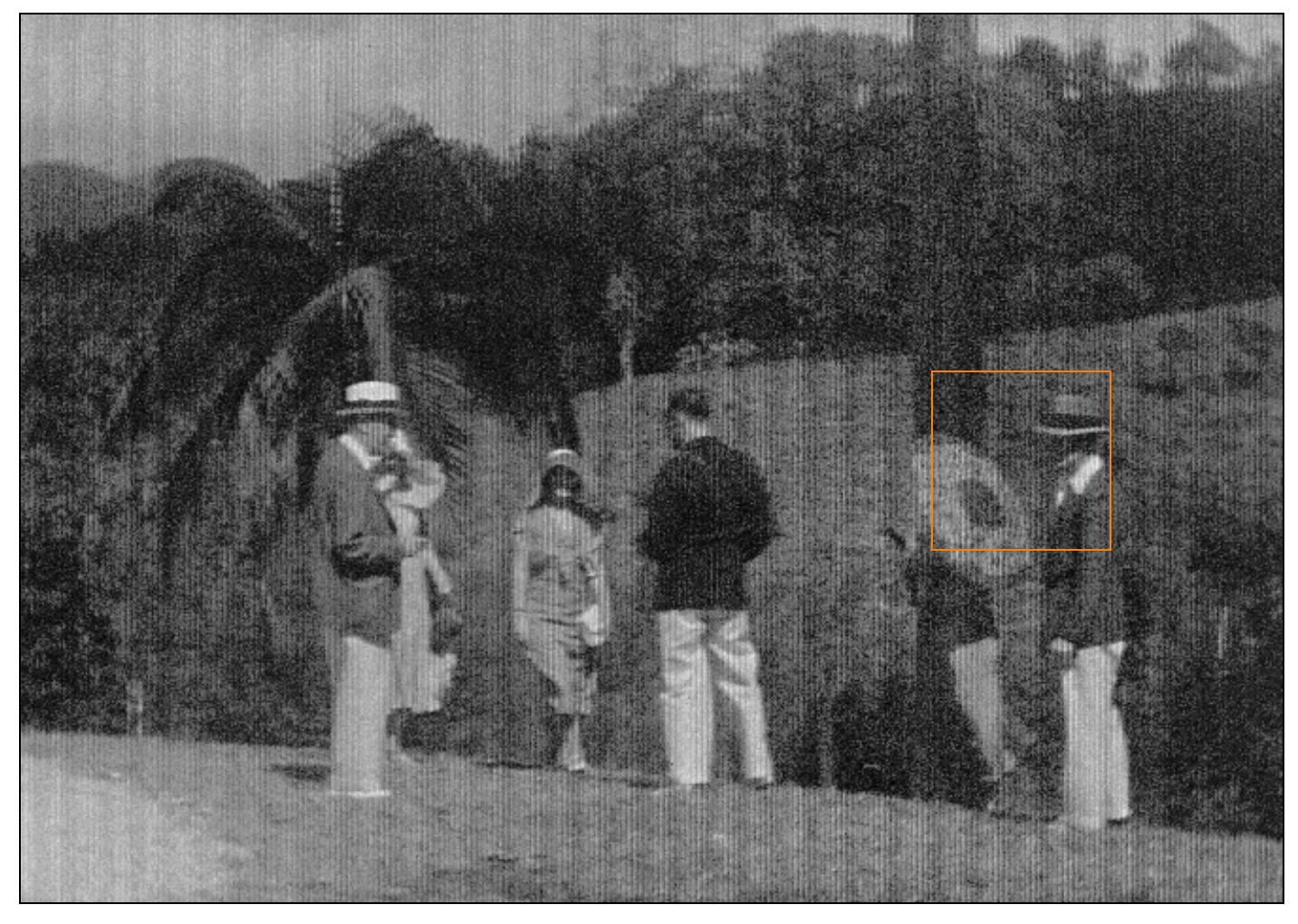}
          \hfill
\includegraphics[height=2.3cm]{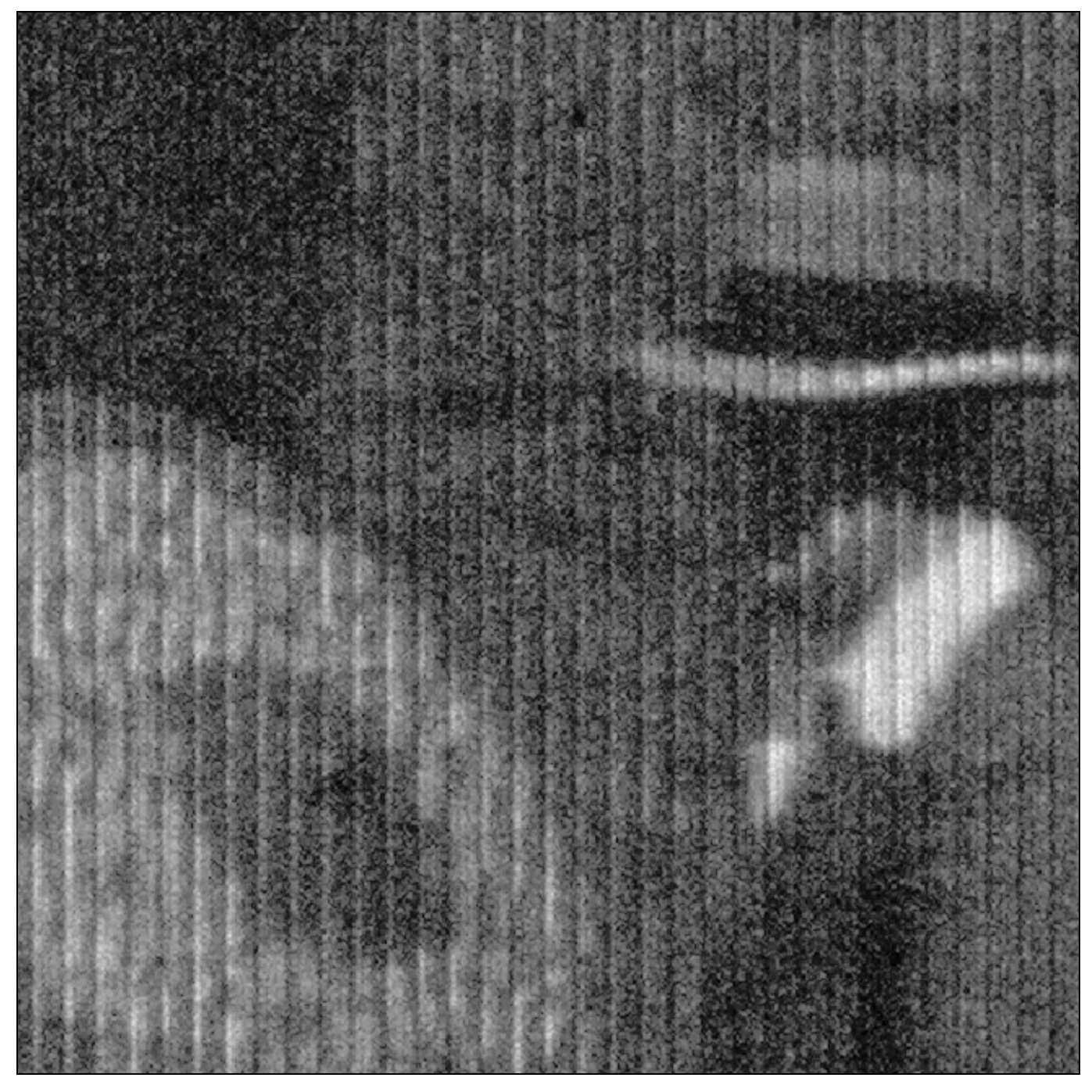}
    \end{subfigure}&
    \begin{subfigure}[c]{0.31\textwidth}
      \includegraphics[height=2.25cm]{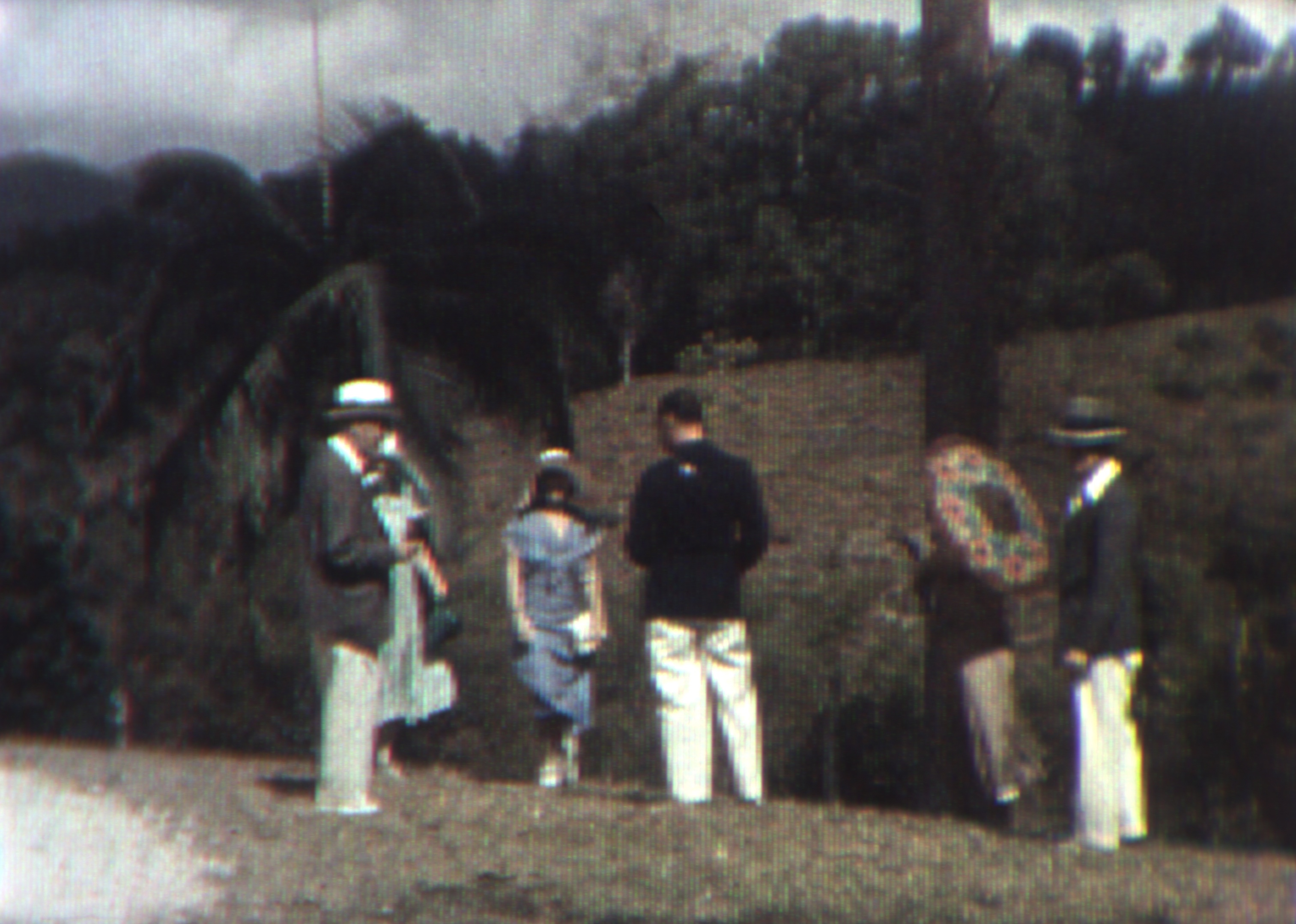}
          \hfill
\includegraphics[height=2.25cm]{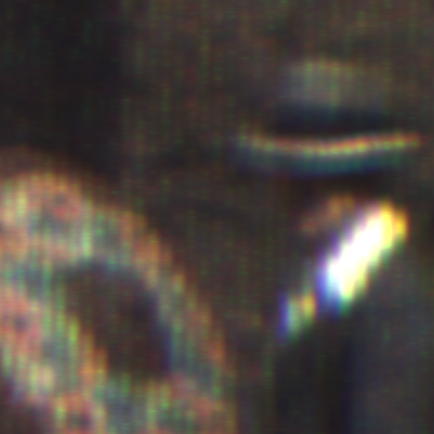}
    \end{subfigure}&
    \begin{subfigure}[c]{0.31\textwidth}
      \includegraphics[height=2.3cm]{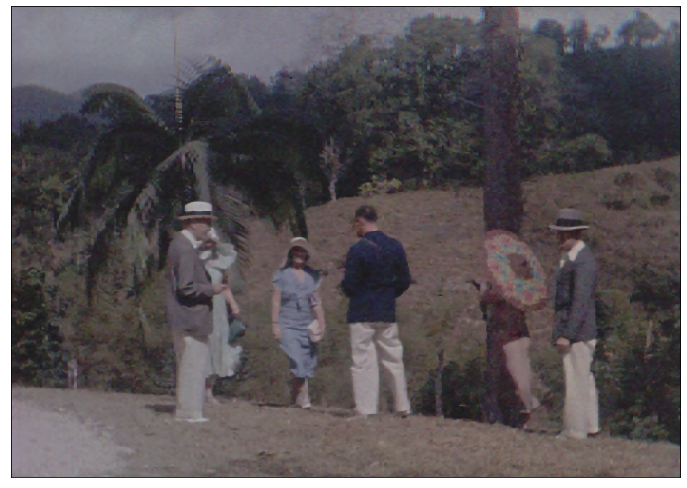}
          \hfill
\includegraphics[height=2.3cm]{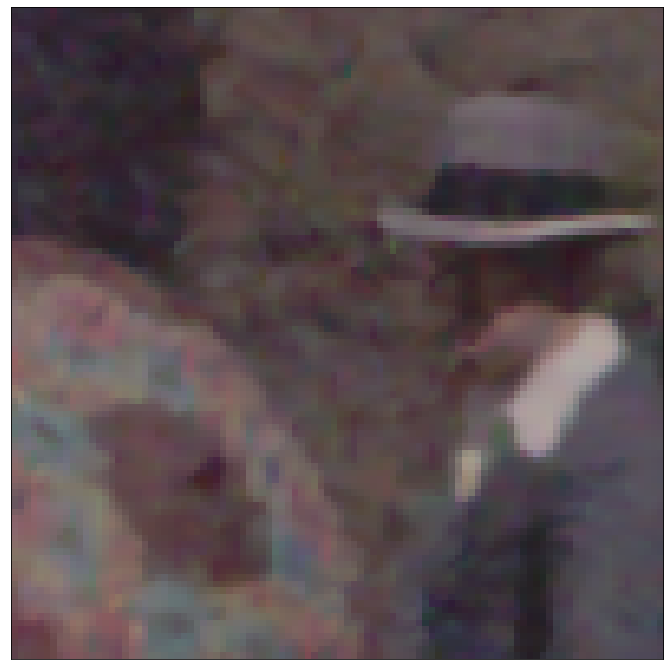}
    \end{subfigure}\\
    
    \begin{subfigure}[c]{0.31\textwidth}
      \includegraphics[height=2.3cm]{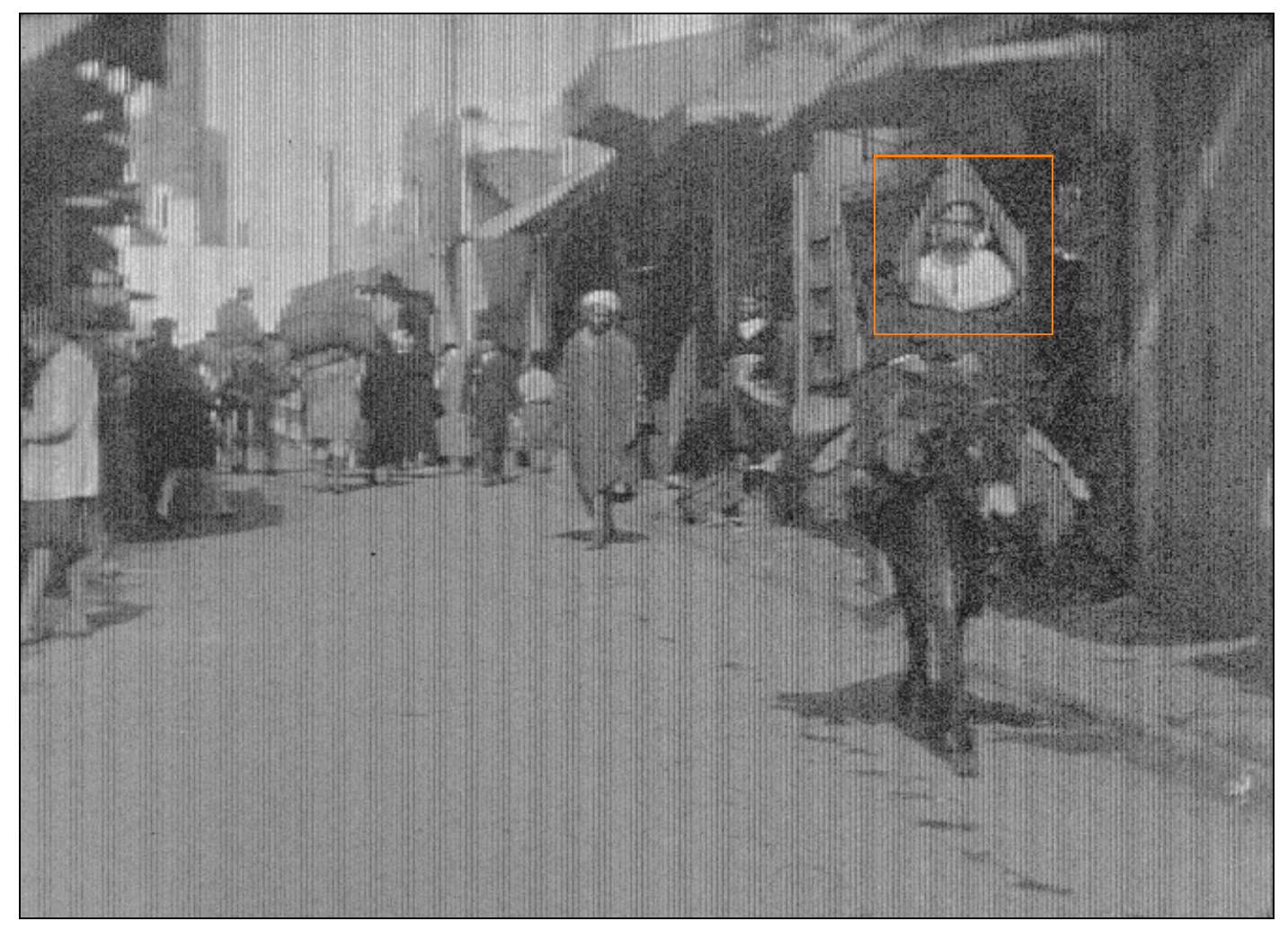}
      \hfill
      \includegraphics[height=2.3cm]{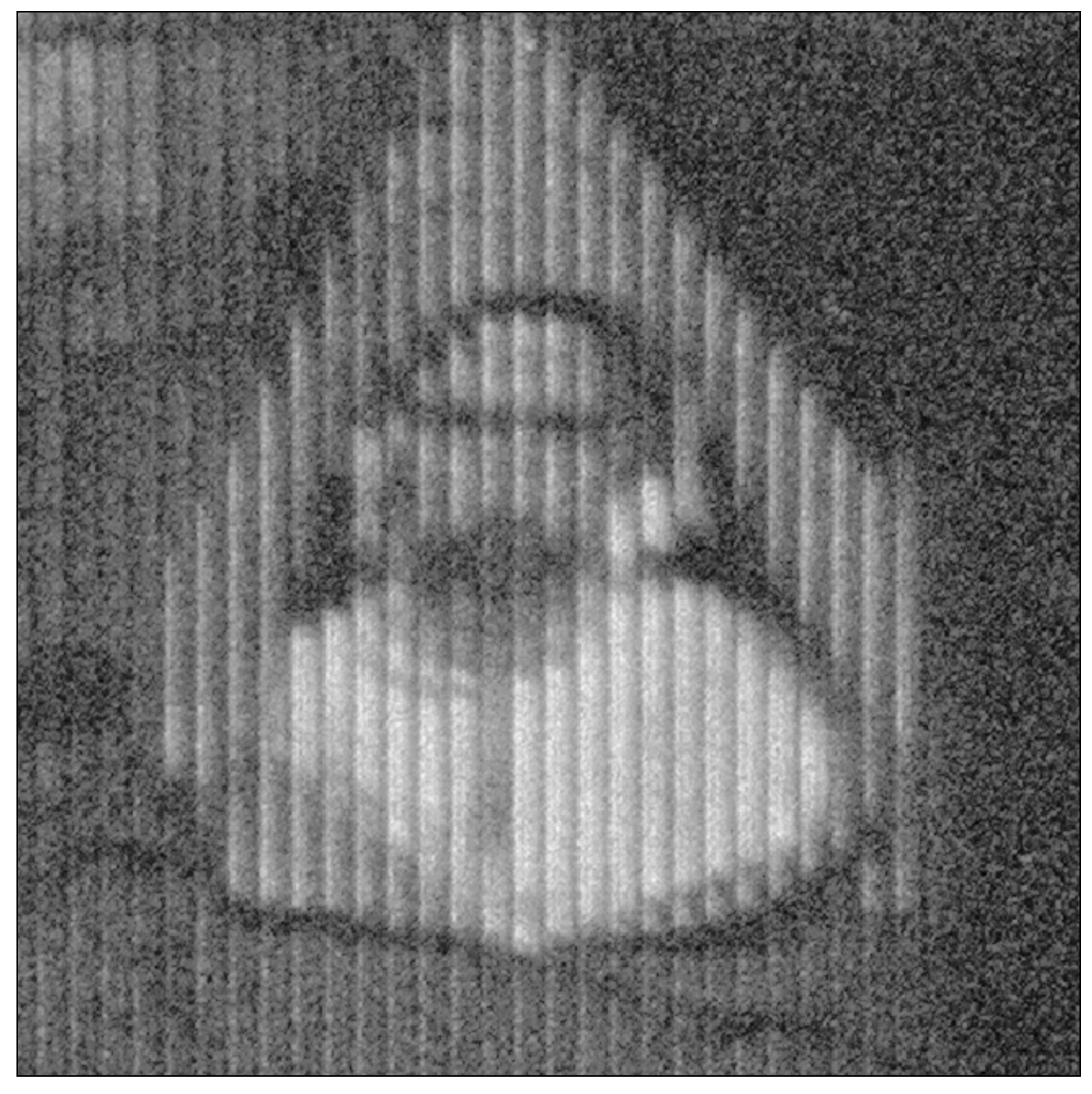}\caption{}
    \end{subfigure}&
    \begin{subfigure}[c]{0.31\textwidth}
      \includegraphics[height=2.25cm]{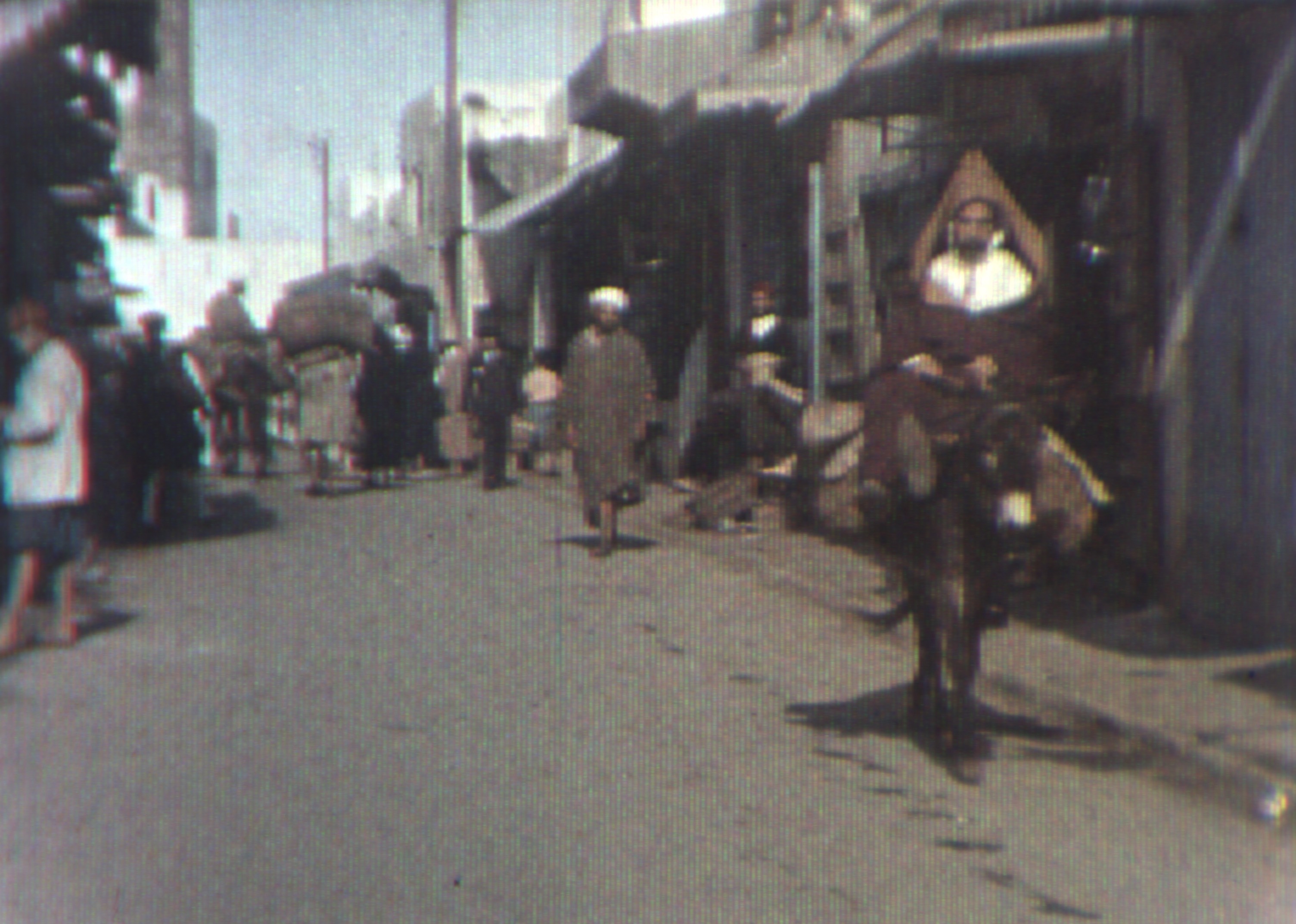}
          \hfill
\includegraphics[height=2.25cm]{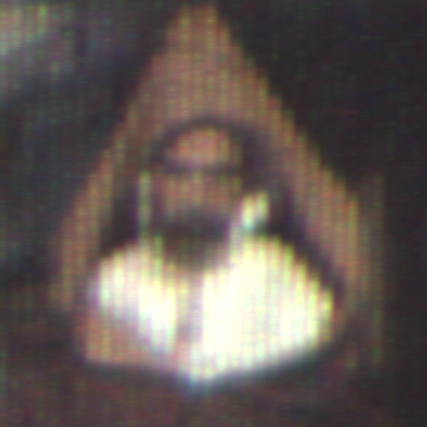}\caption{}
    \end{subfigure}&
    \begin{subfigure}[c]{0.31\textwidth}
      \includegraphics[height=2.3cm]{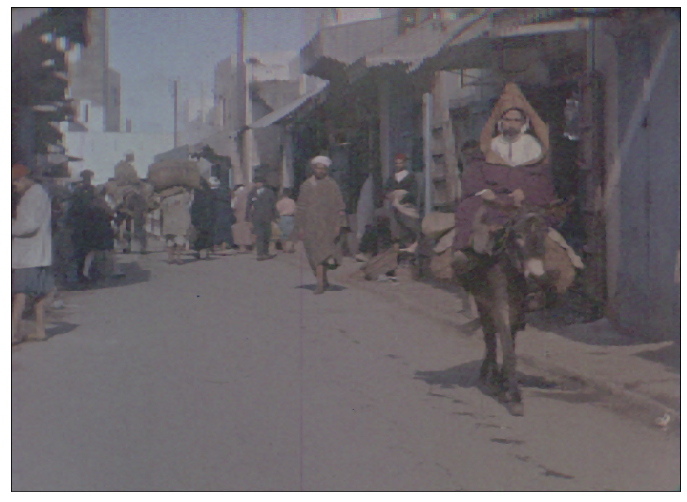}
          \hfill
\includegraphics[height=2.3cm]{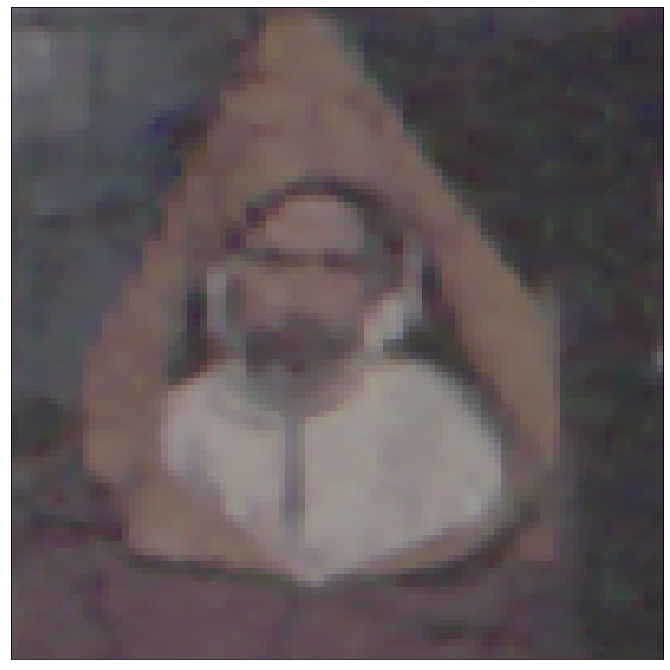}\caption{}
    \end{subfigure}\\
  \end{tabular}    
  \caption{Comparison between (a) input image, (b) analog reconstruction, and (c) color reconstruction with our proposed method.}
  \label{fig:color2}
\end{figure*}

Finally, we provide now a comparison with the original analog color reconstruction. Results can be seen in Fig.~\ref{fig:color2}.
The analog reconstructed color image was obtained by projecting on a digital sensor the color images obtained with the historical projection lens.
First, in terms of color we can observe that the digital proposed method is able to extract truthful colors from the image, although the tone between the analog and the digital colorization is slightly different. Digitally colored images tend to be less saturated with a stronger shade of gray. This can, however, be adjusted by tuning the saturation and brightness of the picture.

The aspect of the images is instead rather different. In the analog reconstructed images the color within each lenticule looks rather constant in the horizontal direction leading to a result that is similar to the output of the doLCE algorithm. The proposed learning-based method is instead able to provide a higher quality colored picture, with sharper details and a more natural look as lenticule boundaries are not visible anymore. This means that the digitally colorized images loose the typical look of the analog projected version. However, the reconstructed colors are truthful, and the details are not hallucinated as the detail level basically matches the grayscale input image. The additional information is actually contained in the lenticular film.

\section{Conclusions}\label{sec:conclusions}
In this work, we proposed a new, learning-based method for color reconstruction of historical, lenticular films. A prerequisite for truthfully reconstructing the colored image is the accurate detection of the lenticule boundaries. Our method is thus composed of two consecutive steps: (\emph{i}) a first deep neural network segments the individual lenticules on a scanned lenticular film. We circumnavigate manual ground truth annotations by training on a dataset where the labels are obtained automatically for artifact-free image samples using doLCE~\cite{Reuteler2014}. (\emph{ii}) After the lenticule locations are refined and the RGB color triplets are extracted, a second deep neural network, trained on a synthetic dataset made of natural images, reconstructs the full color images by learning how to interpolate nearby valid color pixels. The conducted experiments show the validity of the proposed method compared to other existing and baseline methods for lenticular film color reconstruction. Due to the lack of colored groundtruth images, we conducted a user study to evaluate subjectively the proposed colorization technique: the study revealed that the proposed method was largely preferred among the considered methods.
%


%

\section*{Acknowledgments}
The present work leverages the fundamental concepts developed by Rudolf Gschwind and Joakim Reuteler, who initiated the doLCE project. The research would not be possible without the trust and support from Barbara Flueckiger. Special thanks go to the institutions and private collectors who provided the films used in the experiments: Agentur Karl H\"offkes, Stiftung Deutsche Kinemathek, Cinémathèque suisse, Lichtspiel / Kinemathek Bern, Ralf Klee, Haus des Dokumentarfilms Stuttgart.

\appendices
\section{Color Space Estimation}\label{app:color_est}

\begin{figure}
    \centering
    \includegraphics[width=\columnwidth]{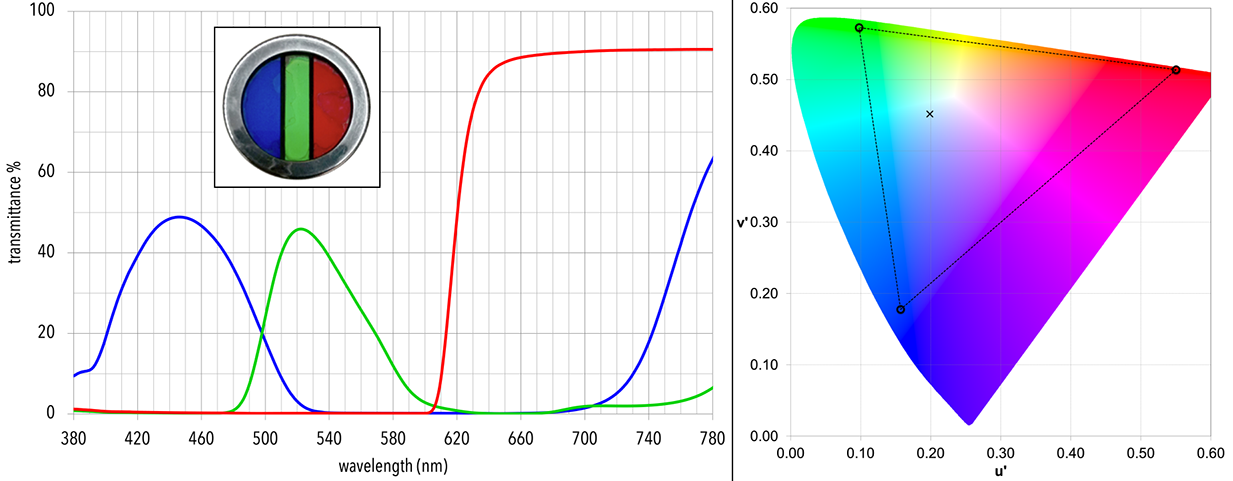}
    \caption{Left: Transmittance spectra of the color filter for lenticular film projection (inset picture). Right: Chromaticity values corresponding to red, green and blue filter's sectors (circles) and corresponding white point (cross).}
    \label{fig:appendix_1}
\end{figure}

During the software color reconstruction the encoded color on the lenticular film must be assigned to the proper RGB space. The \emph{lenticular RGB space} is determined by the transmittances of the color filter originally used in analog projection. 
Transmission spectra (Fig.~\ref{fig:appendix_1} left) were measured with a double-beam spectrophotometer (Shimadzu UV-1800) and colorimetric calculations were performed~\cite{tabulating} considering the 1931 CIE $2^\circ$ standard observer~\cite{colorimetry} and the irradiance spectrum of a typical film projector (Kinoton FP38), which was measured with a spectroradiometer (Konica Minolta CS-2000). The resulting chromaticity values of the color primaries and the corresponding whitepoint are reported in the CIE 1976 UCS diagram (Fig.~\ref{fig:appendix_1} right)~\cite{hunt2011measuring}.

\begin{figure}
    \centering
    \includegraphics[width=\columnwidth]{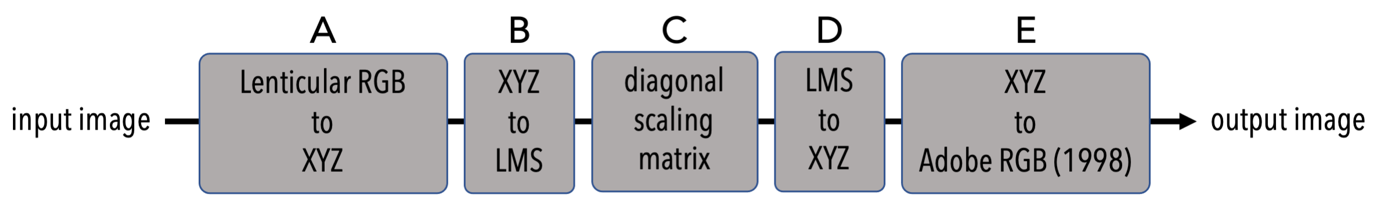}
    \caption{ Five steps of the RGB space conversion with chromatic adaptation transform.}
    \label{fig:appendix_2}
\end{figure}

In order to generate image files that convey the proper color information, the color values of the lenticular RGB space must be converted into a standard RGB space for correct visualization. The whitepoint of the lenticular RGB space ($\text{XYZ} = [0.991, 1, 1.315]$) does not correspond to any standard whitepoint, therefore a chromatic adaptation transform (CAT) is necessary~\cite{moroney2002ciecam02}. In view of the linear von Kries model~\cite{brill1995relation}, the adaptation scaling must be performed at the cone response level (LMS), and therefore the color space conversion requires five steps, which are reported in the diagram of Fig.~\ref{fig:appendix_2}.

\begin{figure}
    \centering
    \includegraphics[width=0.9\columnwidth]{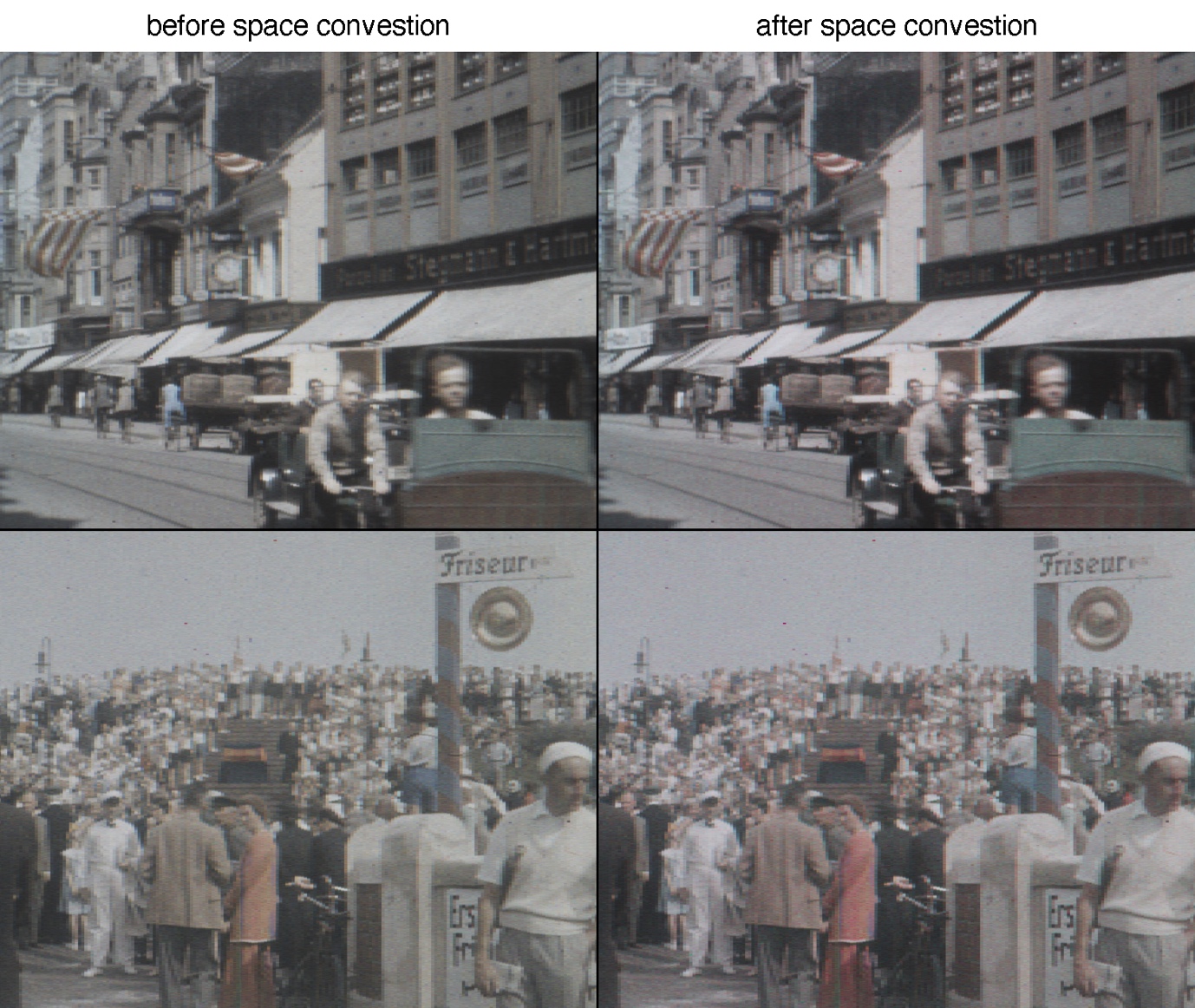}
    \caption{Two examples of color space conversion for correct visualization.}
    \label{fig:appendix_3}
\end{figure}

From the XYZ tristimulus values of the lenticular RGB space, the first $3\times3$ matrix is calculated, which provides the XYZ values of the input RGB image (step-A). The conversion from tristimulus to cone response and back (step-B and step-D) are the $3\times3$ matrices of the CIECAM02 Color Appearance Model~\cite{moroney2002ciecam02}. The chromatic adaptation consists of the scaling factors resulting from the ratios between the LMS values of the destination and provenance whitepoints (step-C). In the present work we chose Adobe RGB (1998) as destination space, whose colorimetric specifications defines the last $3\times 3$ matrix (step-E).
The resulting matrix is the following:
\begin{equation}
\begin{pmatrix}
0.789 & 0.154 & 0.057\\
-0.286 & 1.195 & 0.06 \\
-0.049 & 0.035 & 1.035
\end{pmatrix}
\end{equation}
Two examples of images processed with the space conversion matrix are reported in Fig.~\ref{fig:appendix_3}.

\bibliographystyle{IEEEtran}
\bibliography{bare_jrnl}

\begin{thebibliography}{10}
\providecommand{\url}[1]{#1}
\csname url@samestyle\endcsname
\providecommand{\newblock}{\relax}
\providecommand{\bibinfo}[2]{#2}
\providecommand{\BIBentrySTDinterwordspacing}{\spaceskip=0pt\relax}
\providecommand{\BIBentryALTinterwordstretchfactor}{4}
\providecommand{\BIBentryALTinterwordspacing}{\spaceskip=\fontdimen2\font plus
\BIBentryALTinterwordstretchfactor\fontdimen3\font minus
  \fontdimen4\font\relax}
\providecommand{\BIBforeignlanguage}[2]{{%
\expandafter\ifx\csname l@#1\endcsname\relax
\typeout{** WARNING: IEEEtran.bst: No hyphenation pattern has been}%
\typeout{** loaded for the language `#1'. Using the pattern for}%
\typeout{** the default language instead.}%
\else
\language=\csname l@#1\endcsname
\fi
#2}}
\providecommand{\BIBdecl}{\relax}
\BIBdecl

\bibitem{gordon2013}
M.~Gordon, ``Lenticular spectacles: Kodacolor's fit in the amateur arsenal,''
  \emph{Film History}, 2013.

\bibitem{survey_restoration}
F.~Stanco, G.~Ramponi, and A.~de~Polo, ``Towards the automated restoration of
  old photographic prints: a survey,'' in \emph{The IEEE Region 8 EUROCON.
  Computer as a Tool.}, 2003.

\bibitem{identify_scratches}
R.-C. Chang, Y.-L. Sie, S.-M. Chou, and T.~Shih, ``Photo defect detection for
  image inpainting,'' in \emph{IEEE International Symposium on Multimedia},
  2005.

\bibitem{identify_scratches2}
V.~Bruni and D.~Vitulano, ``A generalized model for scratch detection,''
  \emph{IEEE Transactions on Image Processing}, 2004.

\bibitem{deep_inpaint1}
J.~Yu, Z.~Lin, J.~Yang, X.~Shen \emph{et~al.}, ``Generative image inpainting
  with contextual attention,'' in \emph{Proceedings of the IEEE conference on
  computer vision and pattern recognition}, 2018.

\bibitem{deep_inpaint2}
------, ``Free-form image inpainting with gated convolution,'' in
  \emph{Proceedings of the IEEE/CVF International Conference on Computer
  Vision}, 2019.

\bibitem{back_to_life}
Z.~Wan, B.~Zhang, D.~Chen, P.~Zhang \emph{et~al.}, ``Bringing old photos back
  to life,'' in \emph{Proceedings of the IEEE/CVF Conference on Computer Vision
  and Pattern Recognition}, 2020.

\bibitem{back_to_life_jrnl}
------, ``Old photo restoration via deep latent space translation,''
  \emph{arXiv preprint arXiv:2009.07047}, 2020.

\bibitem{survey_colorization}
S.~Anwar, M.~Tahir, C.~Li, A.~Mian \emph{et~al.}, ``Image colorization: A
  survey and dataset,'' 2020.

\bibitem{deoldify}
A.~Jason, ``Deoldify,'' \url{https://github.com/jantic/DeOldify}, 2018.

\bibitem{old_demosaic}
R.~Kimmel, ``Demosaicing: image reconstruction from color ccd samples,''
  \emph{IEEE Transactions on Image Processing}, 1999.

\bibitem{mosaic_survey}
X.~Li, B.~Gunturk, and L.~Zhang, ``Image demosaicing: A systematic survey,'' in
  \emph{Visual Communications and Image Processing 2008}.\hskip 1em plus 0.5em
  minus 0.4em\relax International Society for Optics and Photonics, 2008.

\bibitem{deep_demosaic_1}
M.~Gharbi, G.~Chaurasia, S.~Paris, and F.~Durand, ``Deep joint demosaicking and
  denoising,'' \emph{ACM Transactions on Graphics}, 2016.

\bibitem{deep_demosaic_2}
G.~Qian, J.~Gu, J.~S. Ren, C.~Dong \emph{et~al.}, ``Trinity of pixel
  enhancement: a joint solution for demosaicking, denoising and
  super-resolution,'' \emph{arXiv preprint arXiv:1905.02538}, 2019.

\bibitem{deep_demosaic_attention}
Y.~Zhang, K.~Li, K.~Li, B.~Zhong \emph{et~al.}, ``Residual non-local attention
  networks for image restoration,'' \emph{arXiv preprint arXiv:1903.10082},
  2019.

\bibitem{Reuteler2014}
J.~Reuteler and R.~Gschwind, ``Die farben des riffelfilms : digitale
  farbrekonstruktion von linsenrasterfilm,'' in \emph{Rundbrief Fotografie},
  2014.

\bibitem{deepdolce_2021}
G.~Trumpy, S.~D'Aronco, J.~D. Wegner, and J.~Reuteler, ``deep-{doLCE}. {A}
  {Deep} {Learning} {Approach} for the {Color} {Reconstruction} of {Digitized}
  {Lenticular} {Film},'' in \emph{Colour {Photography} and {Film} -
  {Conference} {Proceedings}}.\hskip 1em plus 0.5em minus 0.4em\relax
  Associazione Italiana Colore, 2021.

\bibitem{Capstaff}
J.~G. Capstaff, O.~E. Miller, and L.~S. Wilder, ``The projection of lenticular
  color-films,'' \emph{Journal of the Society of Motion Picture Engineers},
  1937.

\bibitem{div2K}
E.~Agustsson and R.~Timofte, ``Ntire 2017 challenge on single image
  super-resolution: Dataset and study,'' in \emph{Proceedings of the IEEE
  Conference on Computer Vision and Pattern Recognition Workshops}, 2017.

\bibitem{unet}
O.~Ronneberger, P.~Fischer, and T.~Brox, ``U-net: Convolutional networks for
  biomedical image segmentation,'' in \emph{International Conference on Medical
  image computing and computer-assisted intervention}.\hskip 1em plus 0.5em
  minus 0.4em\relax Springer, 2015.

\bibitem{resnet}
K.~He, X.~Zhang, S.~Ren, and J.~Sun, ``Deep residual learning for image
  recognition,'' in \emph{Proceedings of the IEEE conference on computer vision
  and pattern recognition}, 2016.

\bibitem{imagenet}
J.~Deng, W.~Dong, R.~Socher, L.-J. Li \emph{et~al.}, ``Imagenet: A large-scale
  hierarchical image database,'' in \emph{IEEE Conference on Computer Vision
  and Pattern Recognition}, 2009.

\bibitem{pytorch}
A.~Paszke, S.~Gross, F.~Massa, A.~Lerer \emph{et~al.}, ``Pytorch: An imperative
  style, high-performance deep learning library,'' in \emph{Advances in Neural
  Information Processing Systems 32}.\hskip 1em plus 0.5em minus 0.4em\relax
  Curran Associates, Inc., 2019.

\bibitem{smp}
P.~Yakubovskiy, ``Segmentation models pytorch,'' \emph{GitHub repository},
  2020.

\bibitem{adam}
D.~P. Kingma and J.~Ba, ``Adam: {A} method for stochastic optimization,'' in
  \emph{3rd International Conference on Learning Representations}, 2015.

\bibitem{lbfgsb}
R.~H. Byrd, P.~Lu, J.~Nocedal, and C.~Zhu, ``A limited memory algorithm for
  bound constrained optimization,'' \emph{SIAM Journal on scientific
  computing}, pp. 1190--1208, 1995.

\bibitem{scipy}
P.~Virtanen, R.~Gommers, T.~E. Oliphant, M.~Haberland \emph{et~al.}, ``{{SciPy}
  1.0: Fundamental Algorithms for Scientific Computing in Python},''
  \emph{Nature Methods}, 2020.

\bibitem{tabulating}
``Recommended practice for tabulating spectral data for use in colour
  computations,'' CIE International Commission on Illumination, Tech. Rep.,
  2005.

\bibitem{colorimetry}
``Colorimetry — part 1: Cie standard colorimetric observers,'' CIE
  International Commission on Illumination, Tech. Rep., 2019.

\bibitem{hunt2011measuring}
R.~W.~G. Hunt and M.~R. Pointer, \emph{Measuring colour}.\hskip 1em plus 0.5em
  minus 0.4em\relax John Wiley \& Sons, 2011.

\bibitem{moroney2002ciecam02}
N.~Moroney, M.~D. Fairchild, R.~W. Hunt, C.~Li \emph{et~al.}, ``The ciecam02
  color appearance model,'' in \emph{Color and Imaging Conference}, 2002.

\bibitem{brill1995relation}
M.~H. Brill, ``The relation between the color of the illuminant and the color
  of the illuminated object,'' \emph{Color Research \& Application}, 1995.

\end{thebibliography}

\end{document}